\documentclass[useAMS,usenatbib]{mn2e}
\usepackage{graphicx}

   \title[The Polar Ring Galaxy A0136-0801.]{Surface Photometry and Metallicity of the Polar Ring Galaxy A0136-0801.}
 
\author[M. Spavone et al.]{M. Spavone$^{1}$\thanks{E-mail: spavone@na.astro.it (MS)}, E. Iodice$^{1}$, M. Arnaboldi$^{2}$\\
   $^{1}$INAF-Osservatorio Astronomico di Capodimonte, via Moiariello 16, I-80131 Napoli, Italy\\
   $^{2}$European Southern Observatory, Karl-Schwarzschild-Strasse 2, D-85748 Garching, Germany}

\begin{document}
\bibliographystyle{mn2e}
   \date{Accepted 2015 March 24.  Received 2015 March 24; in original form 2014 August 22}

\pagerange{\pageref{firstpage}--\pageref{lastpage}} \pubyear{2014}

\maketitle

\label{firstpage}

\begin{abstract}

We present a photometric and spectroscopic study of the polar
  ring galaxy A0136-0801 in order to constrain its formation
  history. Near-Infrared (NIR) and optical imaging data are used to
  extract surface brightness and color profiles of the host galaxy and
  the wide polar structure in A0136-0801. The host galaxy dominates
  the light emission in all bands; the polar structure is more
  luminous in the optical bands and is three times more extended than
  the main spheroid. The average stellar population in the spheroid is
  redder than in the polar structure and we use their (B-K) vs. (J-K)
  colors to constraint the ages of these populations using stellar
  population synthesis models. The inferred ages are 3-5 Gyrs for the
  spheroid and 1-3 Gyrs for the polar structure. We then use long slit
  spectra along the major axis of the polar structure to derive the
  emission line ratios and constrain the oxygen abundance, metallicity
  and star formation rate in this component. We find $12+log(O/H) =
  8.33 \pm 0.43$ and $Z\simeq 0.32 Z_{\odot}$, using emission
    line ratios. These values are
  used, together with the ratio of the baryonic masses of the host
  galaxy and polar structure, to constraint the possible models for
  the formation scenario. We conclude that the tidal accretion of gas
  from a gas rich donor or the disruption of a gas-rich satellite are
  formation mechanisms that may lead to systems with physical
  parameters in agreement with those measured for A0136-0801.

\end{abstract}


\section{Introduction} \label{intro}

In the latest decade, several studies focused on the morphology and
kinematics of polar structures in galaxies, i.e. Polar Ring/Disk
Galaxies (PRGs). These are multi-spin systems, where stars and gas in
the polar structure rotate in a perpendicular plane with respect to
the stars in the central galaxy.  The two decoupled components
  with orthogonal angular momenta are explained as the consequence of
a "second event" in their formation history \citep[see][for a
  review]{Iod2014}. Thus, PRGs are among the ideal galaxies to study
the physics of accretion/interaction mechanisms, the disk formation
and the dark halo shape.
\citet{Whi90} compiled the first PRGs catalogue (PRC), with emphasis on the early-type nature
of the host galaxy (HG), because of its morphology, and on appearance of the polar structure and its relative extension with respect to the mean radius of the HG.
Later studies on the prototype of PRGs, NGC~4650A, showed 
that the polar structure is a disk, rather than a ring \citep[see][]{Arn97,
  Iod02, Gal02, Swa03}. Both in the PRC and in the new SDSS-based
Polar Ring Catalog (SPRC) made by \citet{Moi11}, there are other PRGs
that show a wide disk-like morphology similar to NGC~4650A. They are 
UGC~7576 (PRC A-04), UGC~9796 (PRC A-06), SPRC-27, SPRC-59, and SPRC-69 in the
SPRC, and the galaxy studied in this work A0136-0801 (A-01). Thus, the `PRG' morphological type currently includes both narrow polar rings and wide polar rings/disks. As reviewed by
\citet{Iod2014}, narrow/wide PRGs are characterised by different physical properties in addition to the radial extension of the polar structure.  In particular, in both narrow and wide
PRGs the HG has similar (spheroidal) morphology, colours, and age, but
a different kinematics, independent from the morphological type (narrow or wide
PRGs). Contrary to the HGs, the polar structures in narrow and wide
PRGs have different morphology, baryonic mass, and kinematics, but, on
average, similar oxygen abundance (see Fig.~4 in \citealt{Iod2014}).

Nbody and hydrodynamical simulations try to reproduce the different
systems on the basis of different formation processes.  The up-to-date
formation scenarios proposed for PRGs are: {\it i)} major dissipative
merger; {\it ii)} tidal accretion of material (gas and/or stars) by a
donor; {\it iii)} cold accretion of pristine gas along a filament.  In
the merging scenario, the remnant is a PRG if two disk galaxies merge
on a ``polar'' orbit and have unequal masses \citep{Bou05}. In the
tidal accretion scenario, a polar ring/disk can form around a
pre-existing galaxy by the accretion of gas and stars from the
outskirts of a disk galaxy during a parabolic encounter, or,
alternatively, by the disruption of a dwarf companion galaxy
\citep{Res97, Bou03, Han09}.  Both the two kinds of galaxy
interactions described above (major merging and tidal accretion) are
able to account for several morphologies and kinematics observed for
PRGs, including wide and narrow rings, as well as helical rings and
double rings.  In the framework of disk formation, a polar disk
galaxy, i.e. a galaxy having a polar disk-like structure, forms
through the accretion of pristine gas along a filament \citep{Mac06,
  Bro08}.

The discriminant physical parameters of the formation mechanisms for
PRGs are the baryonic mass, the HG kinematics, and the metallicity in
the polar structure.
These were derived in several works for a sample of PRGs (NGC~4650A,
UGC~7576, UGC~9796; AM2020-504 and VGS31b). The viability of the cold
accretion scenario to form wide polar disk is supported for the first
time \citep{Spav10}. Moreover, it has been confirmed that the tidal
accretion mechanism is able to form both narrow and wide polar rings
\citep{Spav11, Frei12, Spav13}.
 
In this work we present a similar analysis on the wide PRG A0136-0801,
by using both new NIR photometry and spectroscopy, to constrain the
formation history of this galaxy. 

A0136-0801 was discovered by \citet{Sch83} and classified as one of
the best case of \emph{kinematically confirmed polar ring galaxy} (PRC
A-01) by \citet{Whi90}. It is characterized by a wide polar structure
(see Fig.\ref{A0136JR}), which is three times more extended than the
optical radius of the central HG. The H-band images \citep{Iod02b}
showed that the polar structure is less luminous than the central
galaxy, and the bulk of the light is concentrated at smaller
radii. This object was mapped in HI by \citet{Cox04}, who showed that
all HI emission is found within the polar ring, whose outer HI
contours appear to warp away from the poles. The total HI mass for
this object is about $1.6 \times\ 10^{9} M_{\odot}$,
\citet{vanDriel00} also argued that the regular HI distribution and
optical appearance, suggest that the polar structure is quite old and
possibly dynamically stable. Moreover, \citet{Sackett95} found
different HII regions along the polar ring.

\begin{figure*}
\centering
\includegraphics[width=12cm]{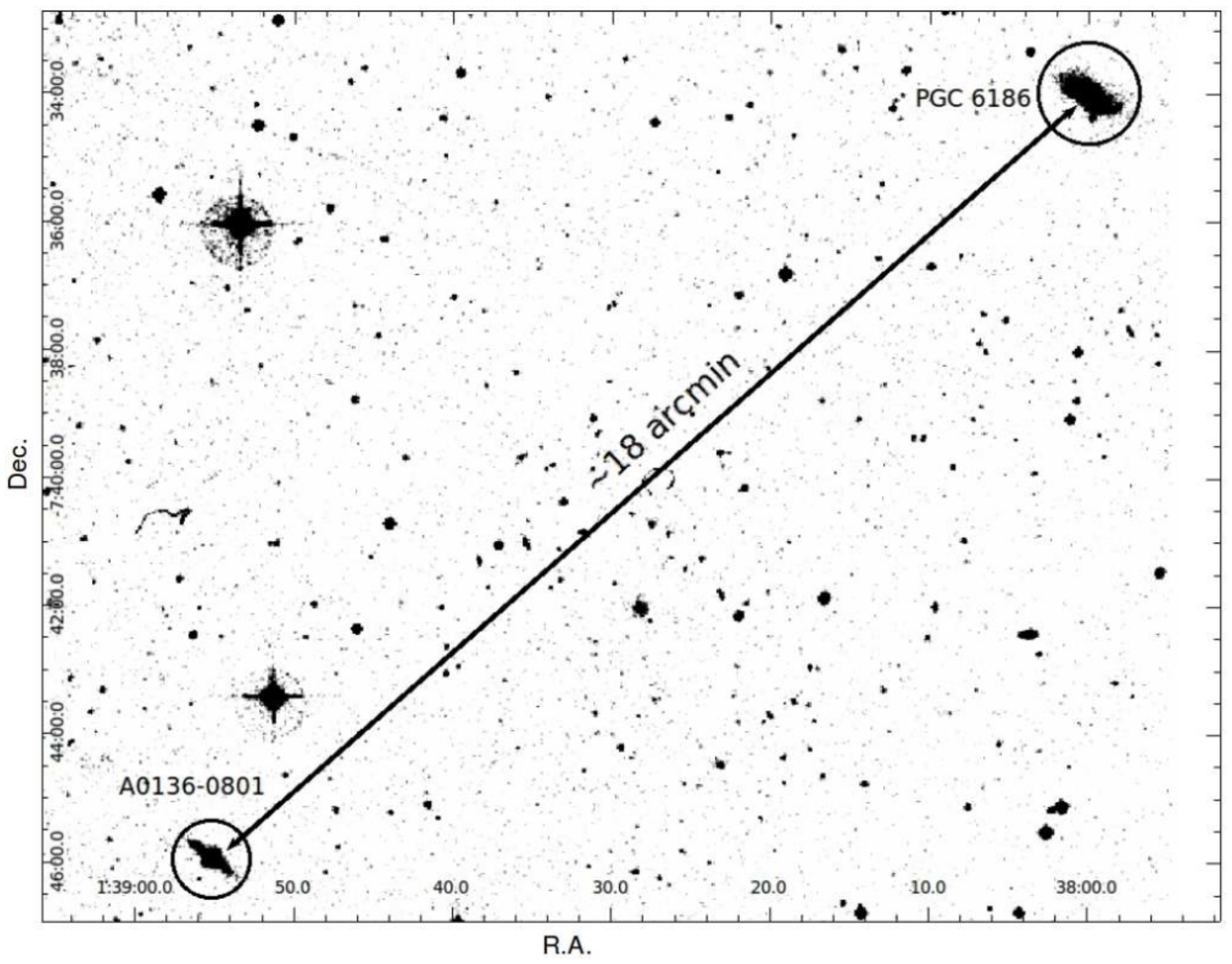}
\caption{SAO-DSS image of the field around A0136-0801, including the nearby galaxy PGC~6186.} \label{field}
\end{figure*}

\begin{figure*}
\includegraphics[width=12cm]{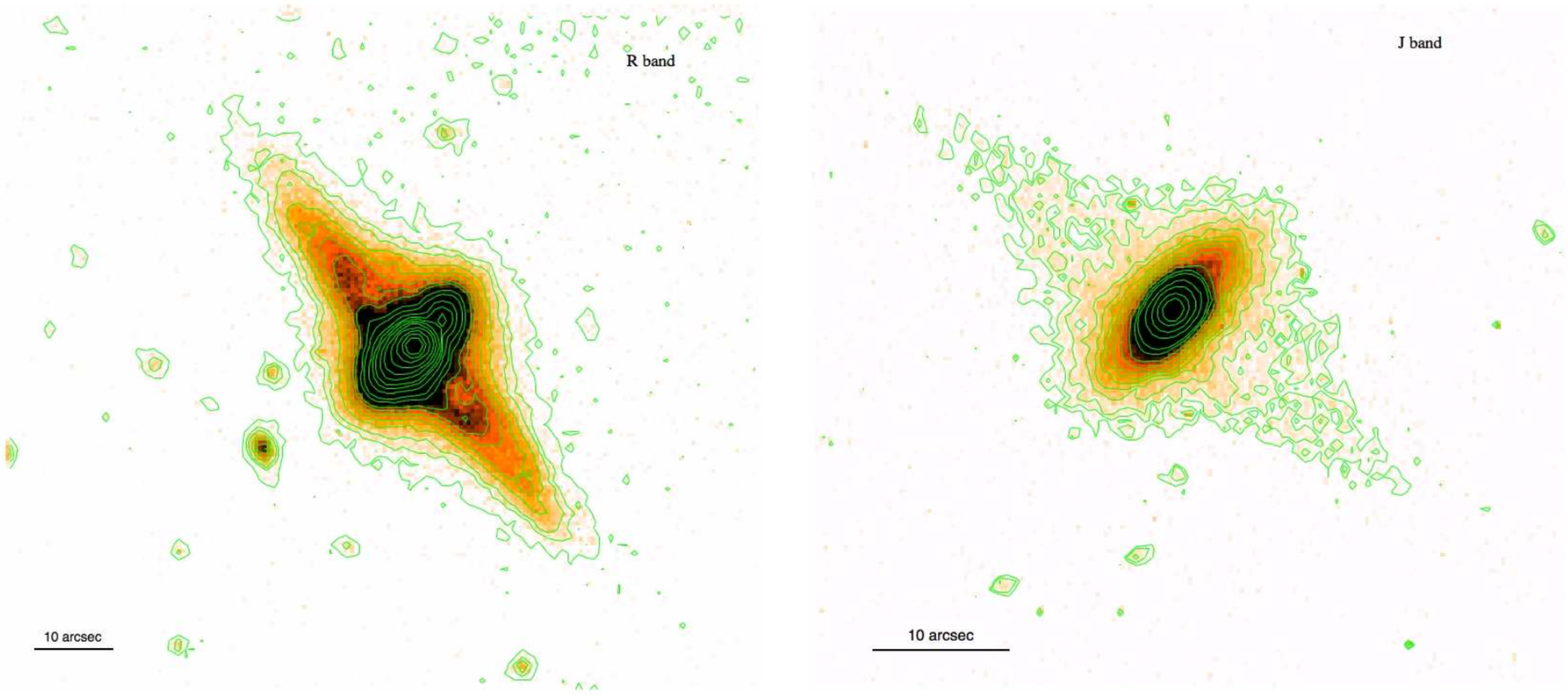}
\includegraphics[width=12cm]{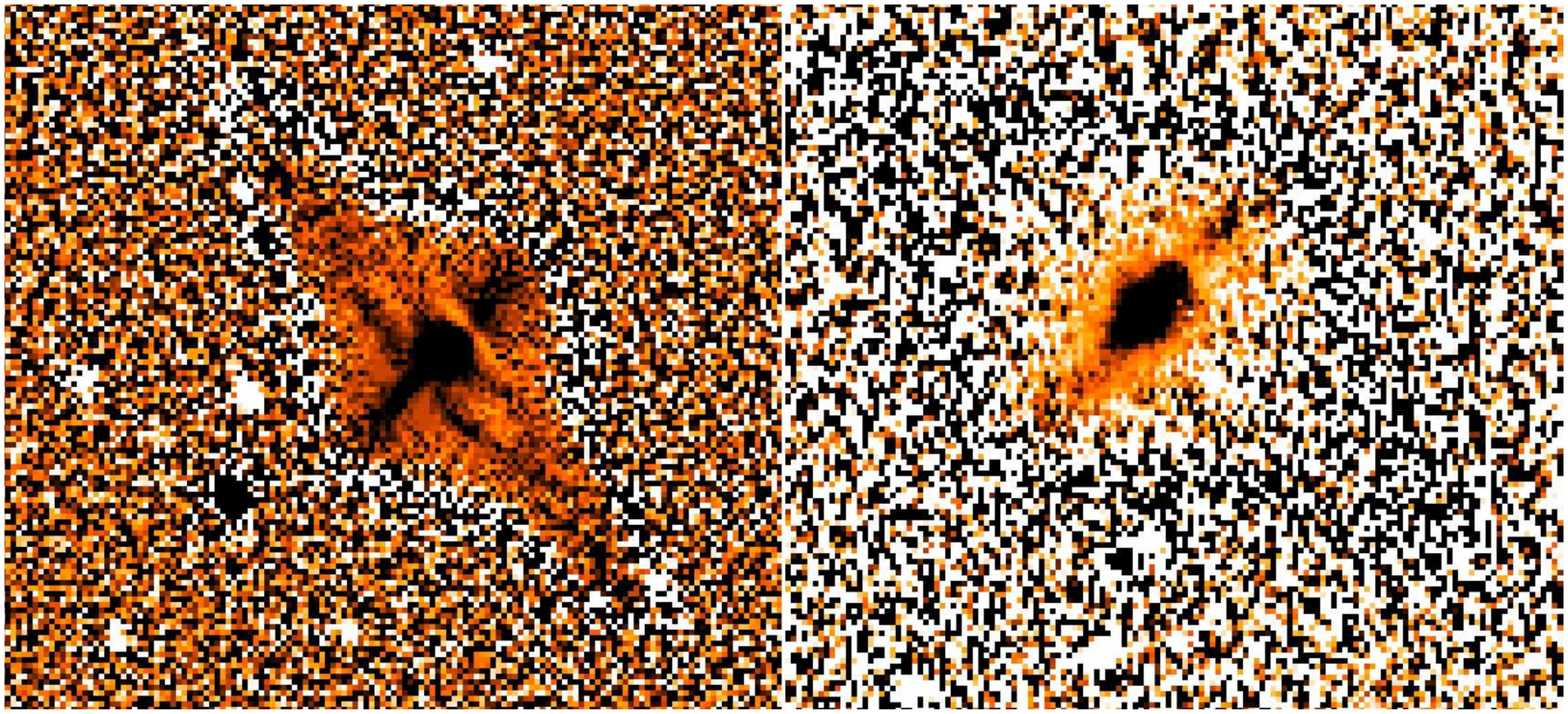}
\caption{Top panels - The polar ring galaxy A0136-0801 in the R band
  (left panel, the image size is $96\times87$~arcsec) and in the J
  band (right panel, the image size is $55\times50$~arcsec). North is
  up and East is on the left. The low and high level of the contours
  (green lines) correspond to $\mu^{R}_{low}=23.87$~mag/arcsec$^2$ and
  $\mu^{R}_{high}=15.52$~mag/arcsec$^2$, in the R band, and
  $\mu^{J}_{low}=20.78$~mag/arcsec$^2$ and
  $\mu^{J}_{high}=16.42$~mag/arcsec$^2$, in the J band.  Bottom panels
  - Enlargement of the high frequency residual images in the R band (left)
  and J band (right).  Darker colors correspond to brighter
  features. The image size is $71'' \times 65''$ for the R band image
  and $41'' \times 37''$ for the J band image.}
\label{A0136JR}
\end{figure*}

In the field around A0136-0801 (Fig.~\ref{field}) there is a larger
galaxy, PGC~6186, which is about 18~arcmin distant from A0136-0801,
and their redshifts differ by 31 km/s \citep{Gal97}.

A0136-0801 has a heliocentric systemic velocity
of $V = 5500$~km~s$^{-1}$, which implies a distance of about 73~Mpc,
based on $H_{0} = 75$~km~s$^{-1}$~Mpc$^{-1}$, which yields an image
scale of 0.3 kpc per arcsec, adopted in this work.  The main properties of A0136-0801 are
listed in Table \ref{global}.

The paper is structured as follow:
in  Sec.~\ref{data} we present the observations and data
reduction; in Sec.~\ref{morph} and Sec.~\ref{phot} we describe the
morphology and structure of the HG and polar structure in A0136-0801,
based on the photometry; in Sec.~\ref{oxy} we give the oxygen
abundances and metallicity of the HII regions in the polar
structure. Results and conclusions are discussed in Sec.~\ref{conc}.

\begin{table*}
\begin{minipage}[t]{\columnwidth}
\caption{Global properties of A0136-0801.}
\label{global}
 \centering
\renewcommand{\footnoterule}{}
\begin{tabular}{lccc}
\hline\hline
Parameter&Value&Ref.\\
\hline
Morphological type&Sc peculiar&NED\footnote{NASA/IPAC Extragalactic Database}\\
R.A. (J2000)           &01h38m55.2s &NED \\
Decl. (J2000)           &-07d45m56s&NED \\
Helio. radial velocity &5500 km/s&NED\\
Redshift         &0.018346 &NED\\
Distance      &73 Mpc     & \\
Diameters&$0.41 \times\ 0.3$ arcmin& \citet{Iod02b}\\
$M(HI)(M_{\odot})$  & $1.6 \times 10^{9}$& \citet{Ric94}\\
$M(H_{2})(M_{\odot})$  & $1.8 \times 10^{9}$& \citet{Gal97}\\
{\it Central galaxy}:&&\\
Absolute magnitude $M_{B}$& -19.26&This work\\
Absolute magnitude $M_{J}$& -21.66&This work\\
Absolute magnitude $M_{K}$& -22.55&This work\\
{\it Polar ring}:&&\\
Absolute magnitude $M_{B}$& -18.49&This work\\
Absolute magnitude $M_{J}$& -19.76&This work\\
Absolute magnitude $M_{K}$& -20.31&This work\\
\hline
\end{tabular}
\end{minipage}
\end{table*}


\section{Observation and data reduction} \label{data}

{\it Near-Infrared data for A0136-0801} - A0136-0801 was in a sample
of PRGs observed in the J and K bands, with the SofI infrared camera
at the ESO-NTT telescope, on December 2002. The camera has a field of
view is $4.92 \times\ 4.92$~arcmin$^{2}$ and a pixel scale of 0.292
arcsec/pixel. A detailed description of the observing strategy and
data reduction is published by \citet{Spav13}. In summary, images were
acquired by adopting an ON-OFF mode where the OFF sky frames were used
to estimate the background level. The total exposure time on the
target in the J and K bands are 360~sec and 1080~sec,
respectively. The average seeing on the final image is FWHM $\simeq$
1.1~arcsec.

The main steps of the data reduction included dark subtraction,
flatfielding correction, sky subtraction and rejection of bad
pixels. The final science frames results from the stacking of the
single pre-reduced exposures. The photometric zero points for that
observing run are $Z_{P}(J) = 23.04 \pm 0.02$~mag/arcsec$^{2}$ and
$Z_{P}(K) = 22.35 \pm 0.02$~mag/arcsec$^{2}$ for the J and K bands,
respecitively.

The calibrated J-band image of A0136-0801 is shown in the right panel of Fig.~\ref{A0136JR}.

{\it Optical data} - Photometric observations for A0136-0801 in the optical bands ($B_{2}$ and
$R_{2}$) adopted for the present work were published by \citet{Godinez07}. They were 
obtained in September 2001 on the 1.5 m telescope
of the Observatorio Astr{\'o}nomico Nacional on Sierra San Pedro
M{\'a}rtir (Baja California, M{\'e}xico), equipped with a SITe1 $1024
\times\ 1024$ CCD detector binned $2 \times\ 2$ to give a pixel scale
of 0.51 arcsec/pixel and a field of view of 4'.3. Reduction of the
CCD frames was performed as described in \citet{Godinez07}. The
photometric calibration was made by using standard stars from the
\citet{Lan83} lists and, after the transformations from the natural
$B_{2}$ and $R_{2}$ to the standard B and R magnitudes (see
\citealt{Godinez07}). We obtained the following photometric zero
points: $Z_{P}(B) = 21.68 \pm 0.01$~mag/arcsec$^{2}$ and $Z_{P}(R) =20.33 \pm 0.01$~mag/arcsec$^{2}$, for the B and R bands respectively.
The calibrated R band image of A0136-0801 is shown in the left panel of Fig.~\ref{A0136JR}. 

{\it Spectroscopic data} - The spectra analyzed in this work were
obtained with the Andalucia Faint Object Spectrograph and
Camera (ALFOSC) at the Nordic Optical Telescope in La Palma, in visitor mode, during the observing run 48-002 (on 28 and 29 October 2013). The adopted slit was $1.3''$ wide and it was aligned
along the major axis of polar structure of A0136-0801, at $P.A. = 135^{\circ}$, in
order to include the most luminous HII regions. The total integration time is 3 hours, with an average seeing of 1.15~arcsec.

We used the grism nr. 4, which coves the wavelength range of
  $3738-6842$\AA, it has a dispersion of 2.3 \AA/pix and a spectral
  resolution $\Delta \lambda \sim$ 10. We aim to detect the the
  red-shifted emission lines of $[OII]\lambda3727$,
  $H_{\gamma}(\lambda4340)$, $[OIII]\lambda4363$,
  $[OIII]\lambda\lambda4959,5007$, $H_{\beta}(\lambda4861)$ and
  $H_{\alpha}(\lambda6563)$ in the polar structure of A0136-0801.

The data reduction and analysis is the same adopted and described by
\citet{Spav10,Spav11, Spav13}, to derive the chemical abundances and
metallicity in PRGs. The main steps includes the CCD
pre-reduction. The wavelength calibration of the spectra was obtained
by comparing spectra of Hg+Ne lamps acquired for each observing
night. The sky spectrum was extracted at the outer edges of the slit,
for $r \ge 30$~arcsec from the galaxy center, where the surface
brightness is fainter than 24~mag/arcsec$^2$, and subtracted off each
row of the two dimensional spectra. The uncertainty on the sky
subtraction is better than $1\%$.  The final median-averaged 2D
spectrum was obtained by co-adding the sky-subtracted frames.

\begin{figure*}
\centering
\includegraphics[width=15cm]{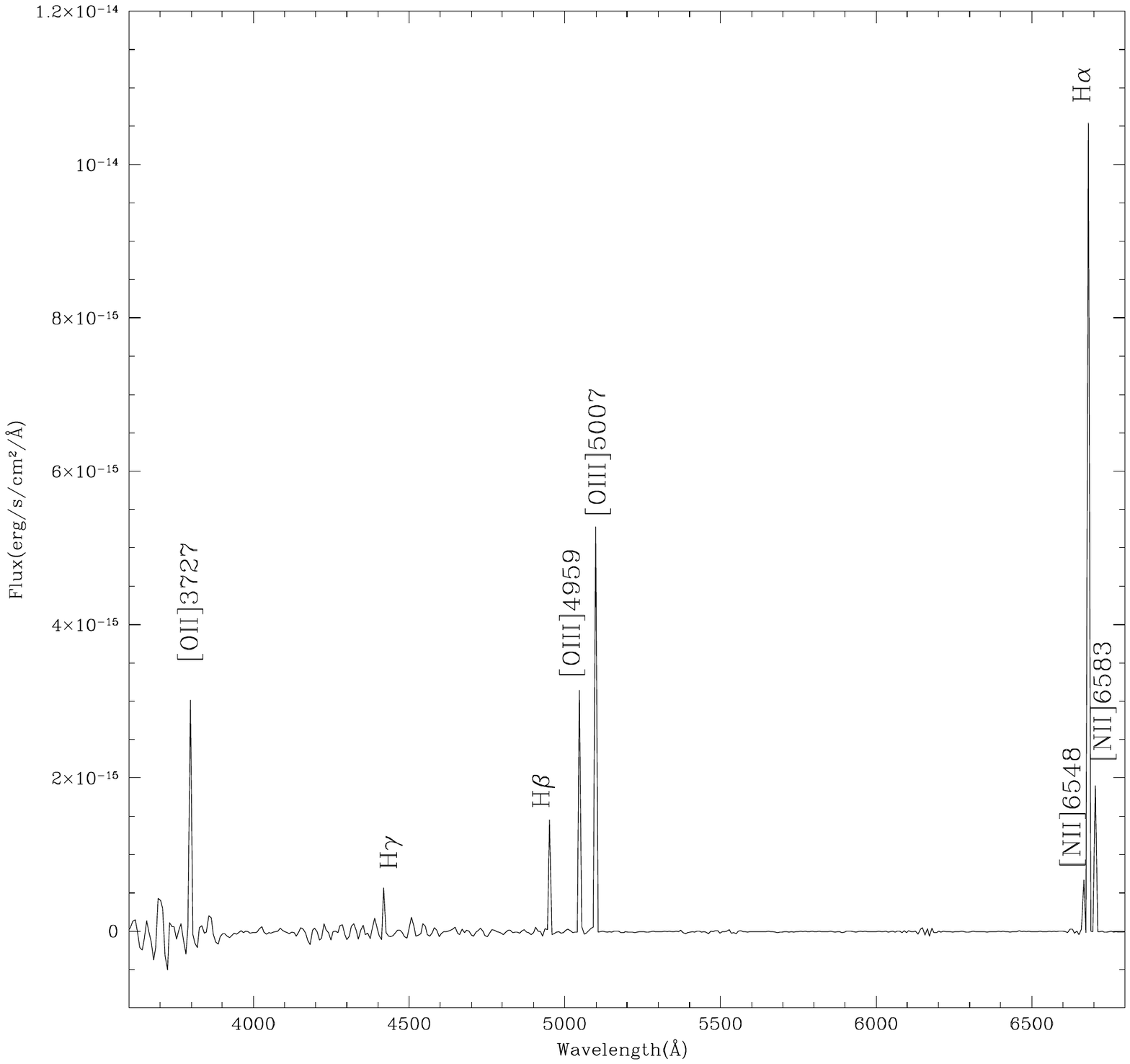}
\caption{Representative 1-D spectrum for the major axis of the polar structure in A0136-0801 obtained by summing up all 1-D spectra along the spatial direction in which emission
  lines had S/N $\ge10$.} \label{spettro}
\end{figure*}

Each 2D spectrum is flux calibrated by using the standard star
HD192281 observed during the first night and HD19445 during the second
night.  The representative 1-D spectrum relative to the major axis of
the polar structure in A0136-0801 is shown in Fig.~\ref{spettro}. It
results by the sum of all 1-D spectra along the spatial direction
having emission lines with S/N $\ge10$.  The integrated flux and the
equivalent width of each emission line was obtained by integrating the
line intensity over the local fitted continuum. The errors estimate on
these quantities was calculated by using the relation published by
\citet{Per03}.

According to \citet {Spav13} [and reference therein], the intensity of
each emission line were corrected for the reddening, which account
both for the absorption intrinsic to the galaxy and to the Milky Way.
Along the polar structure of A0136-0801, the color excess is
$[E(B-V)]$ = $0.24 \pm\ 0.30$, and the optical extinction is
$A(V)=0.76$. This was obtained by comparing the observed and the
intrinsic balmer decrement $H_{\alpha}/H_{\beta}$ measured by summing
up all 1-D spectra along the spatial direction in which the observed
Balmer decrement was greater than the intrinsic one.

By assuming the Cardelli's law \citep{Car89}, $[E(B-V)]$ is used to derive the extinction $A_{\lambda}$. 
The observed and de-reddened mean emission line fluxes relative to $H_{\beta}$ are listed in Table~\ref{fluxes}.

\begin{table*}
\caption{\label{fluxes}Observed and de-reddened mean emission line fluxes relative to $H \beta$.} \centering
\begin{tabular}{lcccc}
\hline\hline
line& $\lambda_{e}$ (\AA) &$\lambda_{obs}$ (\AA) & Observed flux relative to $H \beta$ &De-reddened flux relative to $H \beta$\\
\hline
$H \beta$ & 4861 & 4950 & 1 & 1\\
$H \gamma$ & 4340 & 4420 & 0.4 & 0.45\\
$[OII]$ & 3727& 3795 & 1.99 &2.55\\
$[OIII]$ & 4959&5050 &1.96 & 1.89\\
$[OIII]$ & 5007 &5099 & 3.28 & 3.19\\
$H \alpha$ & 6563 & 6683& 3.72 & 2.89\\
$N[II]$ &6548 &6668 & 0.46& 0.36 \\
$N[II]$ &6583 &6703 & 1.3& 1.02\\
\hline
\end{tabular}
\end{table*}

\section{Host galaxy and polar structure morphology}\label{morph}

Both NIR and optical images show that the central spheroid of
A0136-0801 is the dominant luminous component (Fig.~\ref{A0136JR}, see
also the left panels of Fig.~\ref{modK}). The
polar structure is more clearly visible in the optical bands, while it
is fainter in the J and K bands.  In the optical, it has a diameter of
about 64$''$ ($\sim$ 22 kpc), which is three times more extended than
the optical radius of the central galaxy (20$''$$\sim$ 7 kpc).  This
component hosts several star formation regions and dust that affect
the light distribution \citep{Iod02b}.

To examine the inner structure of the central HG, we have derived the
{\it high frequency residual images} both in the optical and in the
NIR bands. This is the ratio of the original reduced image with a
smoothed\footnote{We used the IRAF task {\small FMEDIAN} to smooth the
  original reduced image} one, where each original pixel value is
replaced with the median value in a rectangular window. The window
size is $7\times 7$~pixels in the R band and $11\times 11$~pixels in
the J band. The un-sharp masked images in these bands are shown in the
bottom panels of Fig.~\ref{A0136JR}.
Both in the optical R-band as well as in the J band, this image shows a
disk-like structure along the major axis of the HG. In the optical
R-band high frequency residual image of A0136-0801, the two arms of
the polar structure are also clearly visible. On the SE side, the
polar structure is in front of the HG, where the dust associated to
this component is obscuring the light coming from the central
galaxy. These features are not detectable in the J-band high frequency
residual image.

\section{Surface Photometry}\label{phot}

In this section we describe the surface photometry and the
two-dimensional model of the light distribution for A0136-0801.  NIR
images, in the J and K bands, where the dust absorption is very low,
are used to study the structure of the central HG in A0136-0801. On
the other hand, since the polar component is much more luminous and
extended in the optical bands, B and R images are mainly used to study
this component. The light and color distributions are
derived for the whole system in both optical and NIR bands.

\subsection{Isophotal analysis}

\begin{figure*}
\centering
\includegraphics[width=8.5cm]{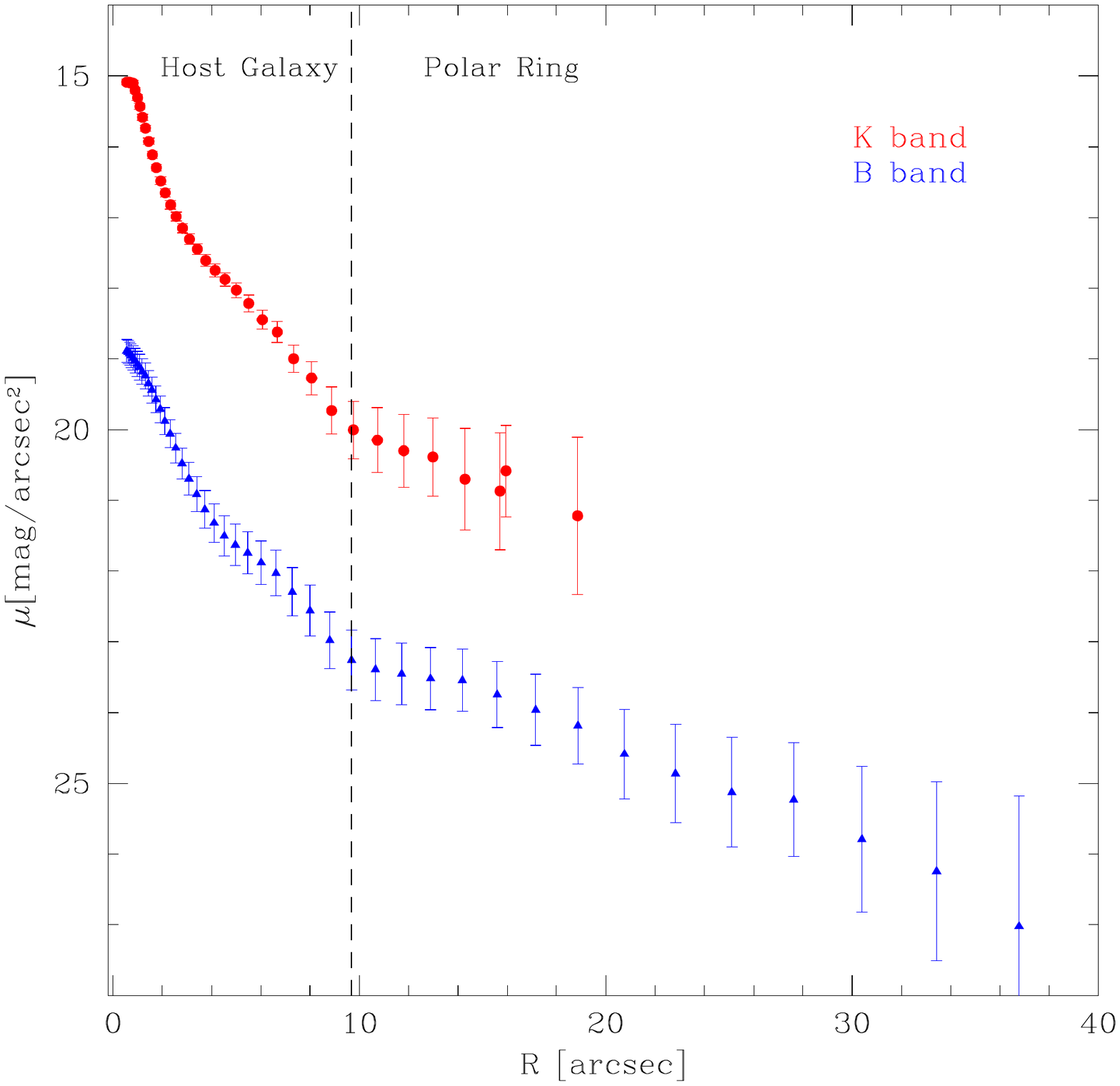}
\includegraphics[width=8.5cm]{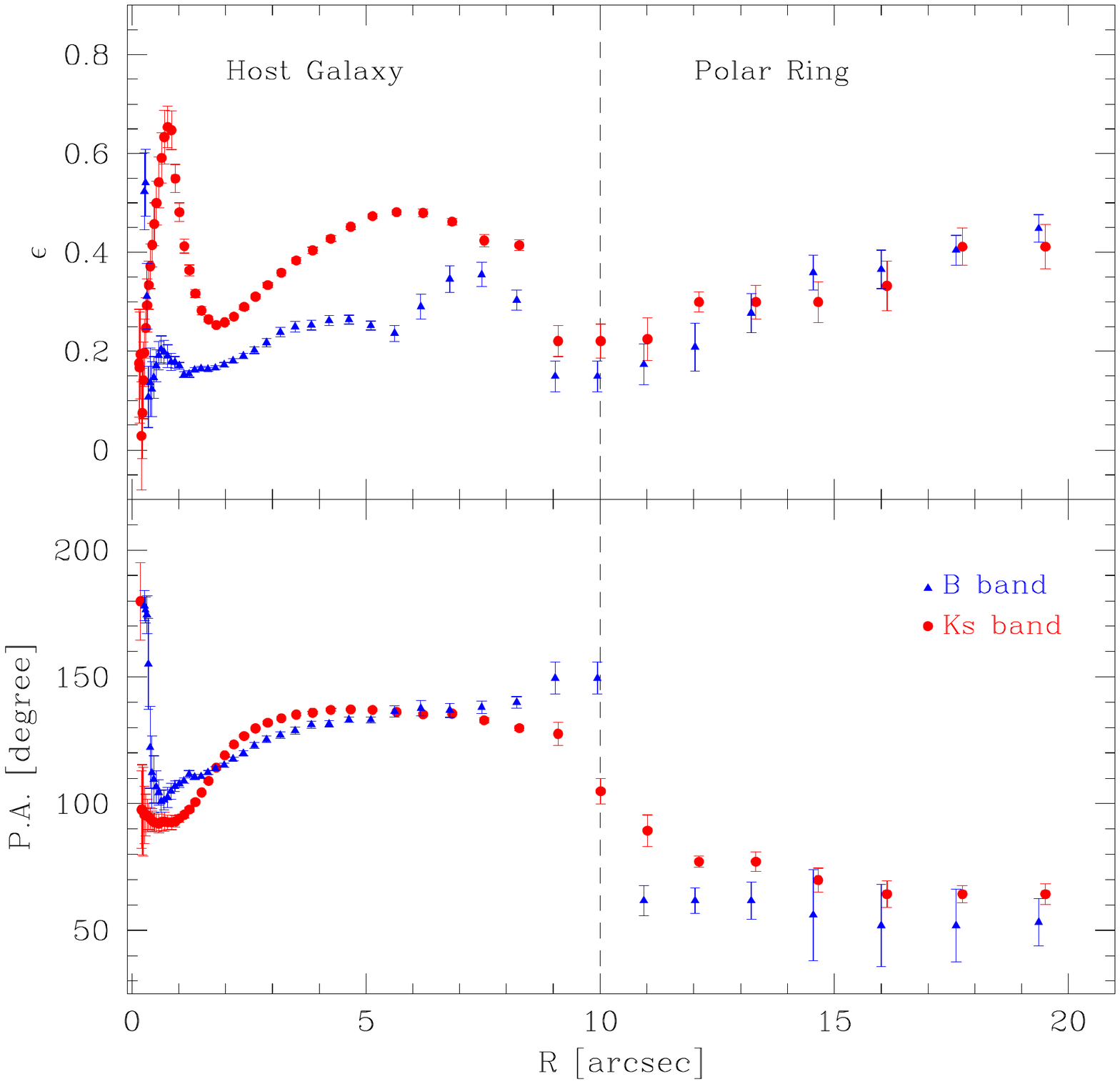}
\caption{Left panel - Azimuthally averaged surface brightness
    profiles as function of $R$, derived by the isophote fit. $R$ is the isophote major
    axis. Data are for the B-band image (triangles, blue points) and K-band (circles, red points). The dashed line delimit the regions where
    the main components of the galaxy structure are located. Right panel - Average profiles of P.A. (bottom panel) and ellipticity (top panel) plotted against the isophote major axis R.} \label{ellipseBK}
\end{figure*}

We used the {\small IRAF-ELLIPSE} task on the whole imaging dataset
to perform the isophotal analysis and to derive the
average surface brightness profiles, Position Angle (P.A.) and
ellipticity ($\epsilon$).  
The limits of the surface photometry presented in this work are 
derived as the distance from the center where the galaxy's light
blends into the background level. These radii
set the surface brightness limit of the optical and NIR photometry. 
Results for the B and K bands are shown in Fig.~\ref{ellipseBK}.
The average surface brightness profiles extend out to 27.77~arcsec
($\sim$8.3~kpc) in the K band and out to 37~arcsec ($\sim$11~kpc) in
the B band (see Fig.~\ref{ellipseBK}, left panel). 
The limiting magnitudes corresponding to the radii given above are $\mu_{K} = 21 \pm 1$~mag~arcsec$^{-2}$ for the
K-band data, and $\mu_B = 27 \pm 2$~mag~arcsec$^{-2}$ for the B-band
image. The error estimates on the above quantities take the
uncertainties on the photometric calibration ($\sim0.01-0.02$~mag) and
sky subtraction ($\sim 0.3\%$~ADU in the B band and $\sim 8\%$~ADU in
the K band) into account.

The nucleus (for $R\le1$~arcsec, $\sim0.3$~kpc) is very bright in the
K band, while light is probably obscured by dust in the optical B band
image. In the K band, the HG  extends out
to 10~arcsec ($\sim3$~kpc). While, in the optical B-band, the
contribution to the light by the polar structure is much more
significant and the semi-major axis of this component is about
27~arcsec ($\sim$8~kpc).

In the range $0\leq R \leq 3$~arcsec ($\leq 1$~kpc), a significant
twisting ($\sim50^{\circ}$) is observed, where the P.A. increases from about 
$ 90^{\circ}$ to $ 140^{\circ}$, and the ellipticity shows a
peak (see Fig.~\ref{ellipseBK}, right panel). This is more pronounced
in the K band ($\epsilon \sim0.65$) than in the B band
($\epsilon \sim0.2$), where the perturbation by dust is stronger.
For $3\leq R \leq 10$~arcsec ($1 \leq R \leq 3$~kpc), in the region
where the HG dominates, the P.A. is almost constant to the value
$\sim145^{\circ}$, and $\epsilon$ increases up to 0.45 in the K image,
and 0.3 in the optical image, and decreases afterwards to $\sim0.2$.
At larger radii, for $10\leq R \leq 20$~arcsec ($3\leq R \leq 6$~kpc),
in the regions of the polar ring, P.A. changes and it is almost
constant in the range of values $\sim50^{\circ}-60^{\circ}$. The
ellipticity shows a linear increase from 0.2 up to 0.4\\

The half-light radii of the whole galaxy are $R_{e}=9.3$~arcsec ($\sim2.8$~kpc) for the
K-band and $R_{e}=14.8$~arcsec ($\sim4.4$~kpc) for the B-band. In
Table~\ref{mag} we give the total integrated magnitudes within two
circular apertures centred on A0136-0801, derived in the K and B
bands. The first aperture is within 14~arcsec, and it was chosen in
order to make the comparison with the integrated magnitudes derived by
the 2MASS data for A0136-0801. The second aperture corresponds to the
outer limit of the surface photometry in the B and K bands given
above.

\begin{table*}
\begin{minipage}[t]{110mm}
\caption{\label{mag}Magnitudes for A0136-0801 in circular apertures.} 
\begin{tabular}{ccccccc}
\hline\hline
Aperture radius& $m_{J}$      & $m_{K}$    & $m_{J} (2MASS)$ & $m_{K} (2MASS)$ & $m_{B}$     & $m_{R}$ \\
               & $\pm 0.02$ & $\pm 0.02$ & $\pm0.02$             & $\pm0.04$              & $\pm 0.01$ & $\pm 0.01$\\
(1)               & (2) & (3) & (4) & (5)  & (6) & (7) \\
\hline
14 & 13.08 & 12.08 & 13.05 & 12.08 & 15.45 & 11.80\\
27.55 &  & 11.58 &  &  &  & \\
36.77 &  &  &  &  & 15.19 & \\
\hline
\end{tabular}
{\em Col.~1}: Radius  of the circular aperture in arcsec.  {\em Col.~2} and {\em Col.~3}: Integrated magnitudes in the J and K bands from the SOFI data. {\em Col.~4} and {\em Col.~5}: Integrated magnitudes in the J and K bands from the 2MASS data. {\em Col.~6} and {\em Col.~7}: Integrated magnitudes in the optical B and R bands.
\end{minipage}
\end{table*}

\subsection{2-Dimensional model of the light distribution}\label{2D}

A0136-0801 has two main components, the central spheroid (the HG),
with a bright nucleus, and the polar structure, which resembles a ring
(see Fig.~\ref{A0136JR}).  In order to measure the structural
parameters of the main galaxy components we adopted the following
approach. In the K-band, which is less perturbed by dust, we obtained
the best 2-dimensional (2D) model of the light distribution for the
central HG, which is the dominant component at this wavelength, having
masked the regions where the polar ring is still detectable. For the
optical B-band image, both the HG and the polar ring light has been
modeled.

The light distribution in the HG is modelled by a Sersic law \citep{Ser68}, to reproduce the bright nucleus,
\begin{equation}
\label{sersic}
\mu (R) = \mu_e + k(n) \left[ \left( \frac{R}{r_e}\right)^{1/n} -1\right]
\end{equation}

plus an exponential law, to account for the disk-like structure detected in the high frequency residual images of A0136-0801 (see Fig.~\ref{A0136JR}), given by 
\begin{equation}
\mu(R)= \mu_{0} + 1.086 \times R/r_{h}
\end{equation}

where $R$ is the galactocentric distance, $r_e$ and $\mu_e$ are the
effective radius and effective surface brightness of the nucleus, and
$k(n)=2.17 n - 0.355$, $\mu_{0}$ and $r_{h}$ are the central surface
brightness and scale length of the disk.  In the B-band image, the 2D
model accounts for both the light distribution of the HG (as
described above) and polar structure. For this second component
we used an exponential law, and derived the central surface brightness
$\mu^{PR}_{0}$ and scale length $r^{PR}_{h}$ of the polar ring.

\begin{figure*}
\centering
\includegraphics[width=17cm]{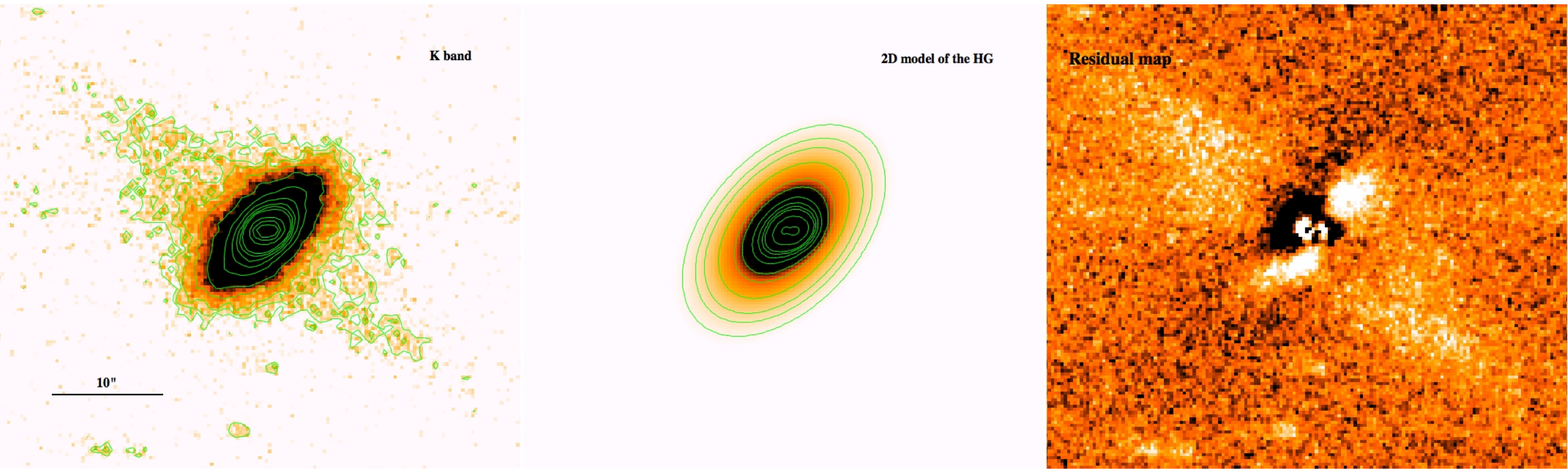}
\includegraphics[width=17cm]{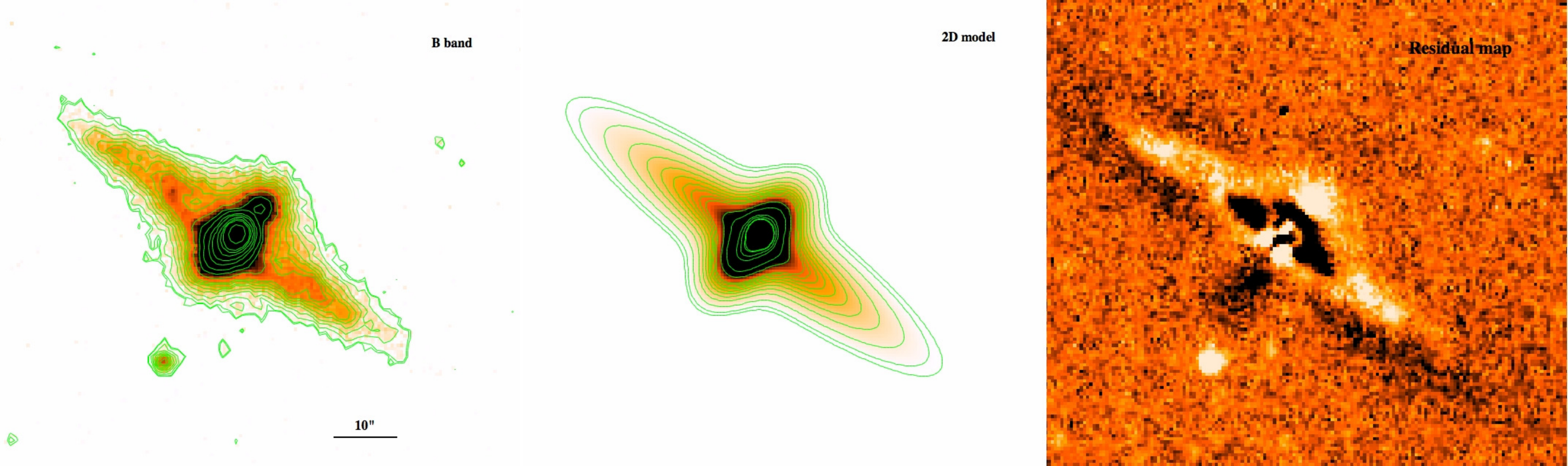}
\caption{Top: 2D model of A0136-0801 in the K band. Left panel - K
  band image of A0136-0801. Middle panel - 2D model of the HG. Right
  panel - Residual of the subtraction of the model to the K band
  image. Bottom: 2D model of A0136-0801 in the B band. Left panel - B band image of A0136-0801. Middle panel - B band model of the whole system (HG and polar ring). Right panel - Residual of the subtraction of the model to the B band image.} \label{modK}
\end{figure*}

The maximum symmetric 2D model is made by using the GALFIT package
\citep{Peng02}, where all the structural parameters listed above,
including also the total magnitudes, axial ratios and P.A.s are left
free. A summary of the best-fit parameters for each component is
listed in Table~\ref{galfit} and the results are shown in
Fig.~\ref{modK}.

\begin{figure*}
\centering
\includegraphics[width=8.5cm]{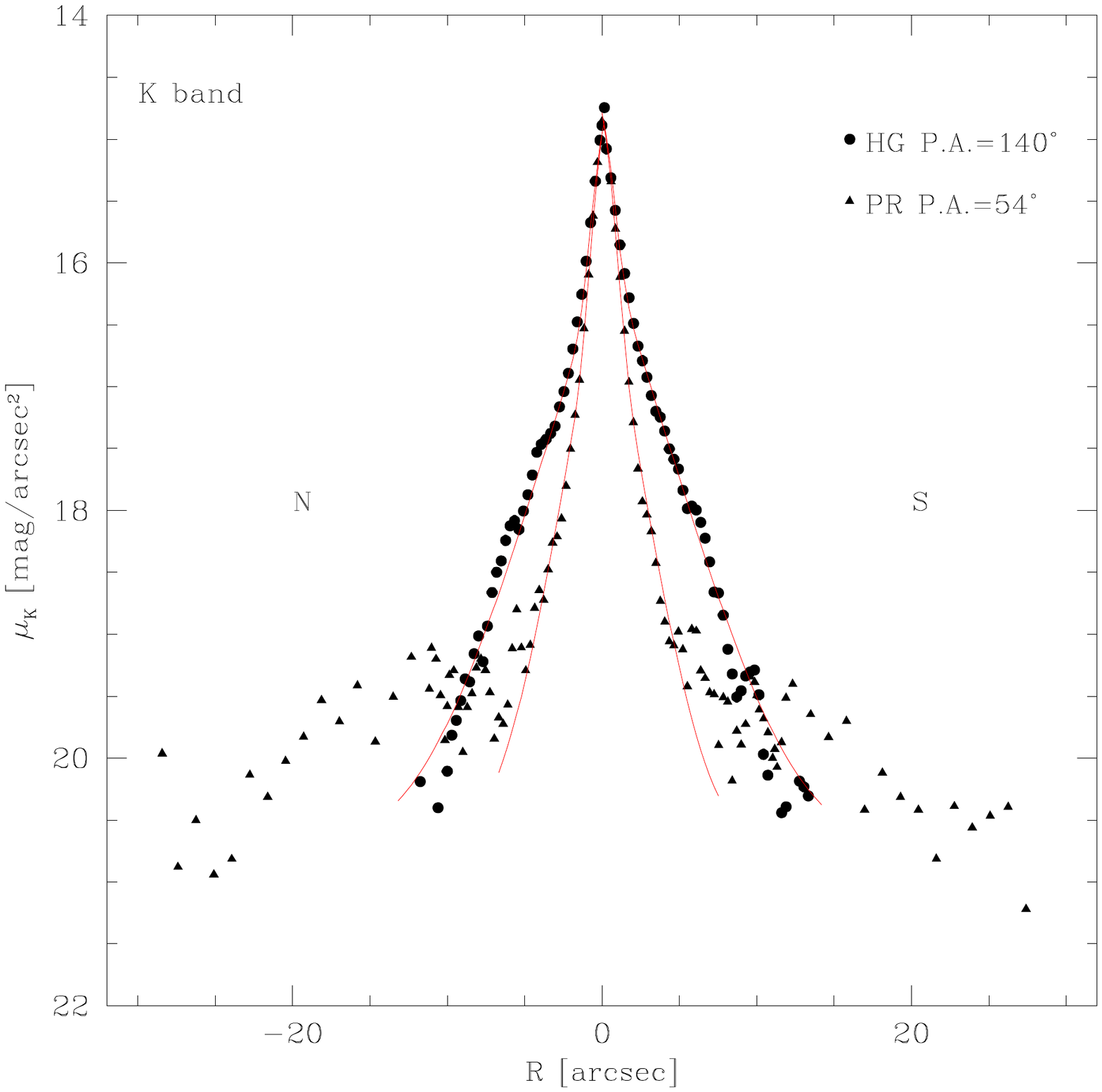}
\includegraphics[width=8.5cm]{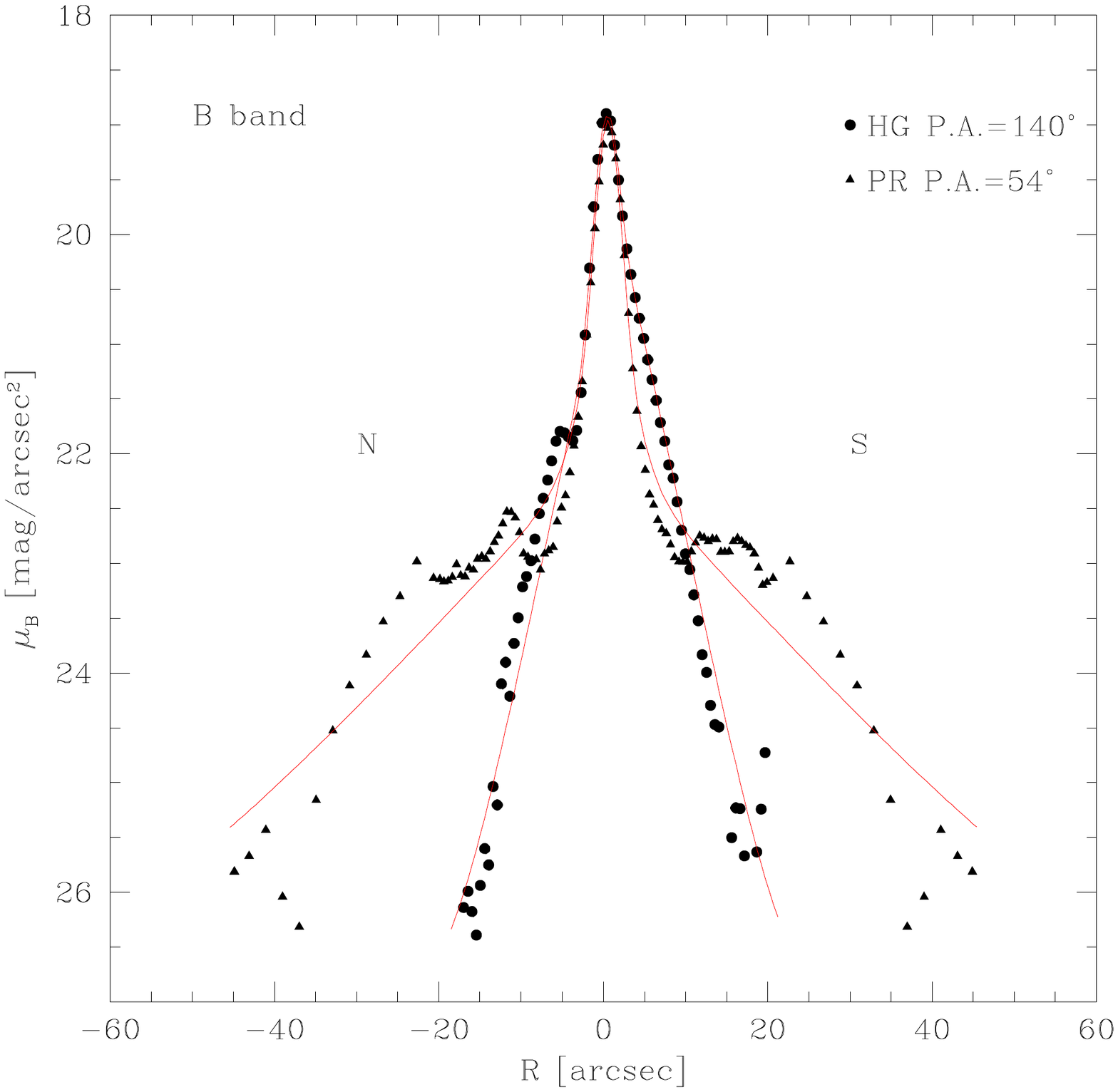}
\caption{Comparison between the observed light profile along the HG major axis (circles) and polar ring  (triangles) and those derived by
  the 2D model (solid red lines) in the B and K bands.} \label{profmod}
\end{figure*}

The morphology of the HG is well reproduced by the 2D model in the K
band (see top-middle panel of Fig.~\ref{modK}). The best fit to the
light distribution is obtained by the superposition of two component:
a very concentrated bulge-like structure, with an effective radius of
1.12~arcsec ($\sim0.34$~kpc) and and exponential disk, with a scale
length of about 3~arcscec ($\sim0.9$~kpc), see Table~\ref{galfit}. The
two components have similar flattening, but different P.A., consistent
with the results of the isophote analysis (see Fig.~\ref{ellipseBK}).
A twisted bulge-like component, in the central regions of the HG, is a
real feature as suggested by the isophotal contours shown in top-right
panel of Fig.~\ref{modK}. This misaligned structure is well reproduced
by the 2D model in the K band (see top-middle panel of
Fig.~\ref{modK}). The residual map
(see top-right panel of Fig.~\ref{modK}), obtained by subtracting the
2D model from the original image, shows a nuclear sub-structure having
an ``s-shape'', elongated towards the polar direction, and two regions
on both sides along the major axis of the HG where the galaxy is
brighter than the model. The 2D residuals also clearly points out the
low emission by the polar ring, which has been masked in the fitted
image.  Even if the optical B-band image is much more affected by the
dust absorption, the 2D model of the light distribution reproduces
consistently the morphology of both the HG and polar ring (see
bottom-middle panel of Fig.~\ref{modK}). The scale radii of the two
components in the HG (bulge-like and disk) are comparable with those
derived by the fit of the K-band image (see Table~\ref{galfit}). One
interesting result is that the P.A. of the bulge-like nucleus is
consistent with the P.A. found for the polar ring. The residual map of
the B-band 2D model (see bottom-right panel of Fig.~\ref{modK}) also
shows the luminous ``s-shape'' structure in the center, as already found
in the residual map of the K-band model. The optical residual map
shows several bright bumps along the polar ring, which are due to the
star forming regions, and the loci where the two arms of the polar
ring cross the HG, one on the SE side (behind the galaxy) and the
other one on the NW side that passes in front of the HG, where
the dust absorption is stronger.  The comparison between the observed
and fitted light profiles, in the B and in the K bands, along the HG
major axis and along the polar ring, are shown in
Fig.~\ref{profmod}. The empirical laws for the main components
observed in A0136-0801 are a good description of the average light
distribution.

The implications of this analysis on the structure of A0136-0801, in
particular the nature and the origin of the sub-structures identified
in the residual maps, will be discussed in Sec.~\ref{conc}.

\begin{table*}
\begin{minipage}{180mm}
\caption{Structural parameters for the 2D model of the light  distribution of A0136-0801 in the B and K bands.} 
\label{galfit}
\begin{tabular}{l c c c c c c c c c}
\hline\hline
Component & Model & $m_{tot}$ & $\mu_{e}$ & $\mu_{0}$ & $r_e$ & $r_h$ & $n$ & P.A.& $\epsilon$ \\
  &  & mag & mag arcsec$^{-2}$ & mag arcsec$^{-2}$ & arcsec  & arcsec &  & degree & \\  
(1)  & (2) & (3) & (4) & (5) & (6)  & (7) & (8) & (9) & (10) \\  
\hline\hline                    
 & & & &{\it K band} & & & &  & \\
\hline
HG - Bulge & Sersic & $13.60\pm0.01$ & $ 17.23\pm0.03$ &  & $1.12 \pm 0.01$ &  & $0.44 \pm 0.01$ & $95.5 \pm 0.2$ & $0.33 \pm 0.01$\\ 
HG - Disk  &  exp      &$12.20\pm0.01$ &  & $ 16.24\pm0.04$ & & $2.56 \pm 0.04$ &  & $135.7\pm0.2$ & $0.43\pm0.01$  \\ 
\hline                 
 & & & & {\it B band} & & & & & \\
\hline
HG - Bulge & Sersic & $16.96\pm0.01$ & $ 20.85\pm0.02$ &  & $1.26 \pm 0.01$ &  & $0.50 \pm 0.01$ & $52 \pm 5$ & $0.03 \pm 0.01$\\ 
HG - Disk  &  exp      &$16.18\pm0.01$ &  & $ 20.56\pm0.02$ & & $3.00 \pm 0.02$ &  & $139.6\pm0.2$ & $0.50\pm0.01$  \\ 
PR             & exp       &$15.99\pm0.01$ &  & $23.60\pm0.04$ & & $13.3 \pm 0.2$ &  & $54.3\pm0.1$ & $0.78\pm0.02$\\ 
\hline\hline
\end{tabular}
{\em Col.1}: Different components observed in A0136-0801. {\em Col.2}: Empirical law adopted to fit the light distribution for each component. {\em Col.3}: Total magnitude corresponding to each component. {\em Col.4 - Col.8}: Structural parameters that characterise each empirical law (i.e. effective surface brightness $\mu_e$, effective radius  $r_e$ and $n$-exponent of the Sersic law, and central surface brightness $\mu_0$ and scalelength $r_h$ for the exponential law). {\em Col.9} and {\em Col.10}: Average Position Angle and ellipticity of the isophote.
\end{minipage}
\end{table*}

\section{Color distribution and integrated magnitudes}\label{colors}

\begin{figure*}
\includegraphics[width=8cm]{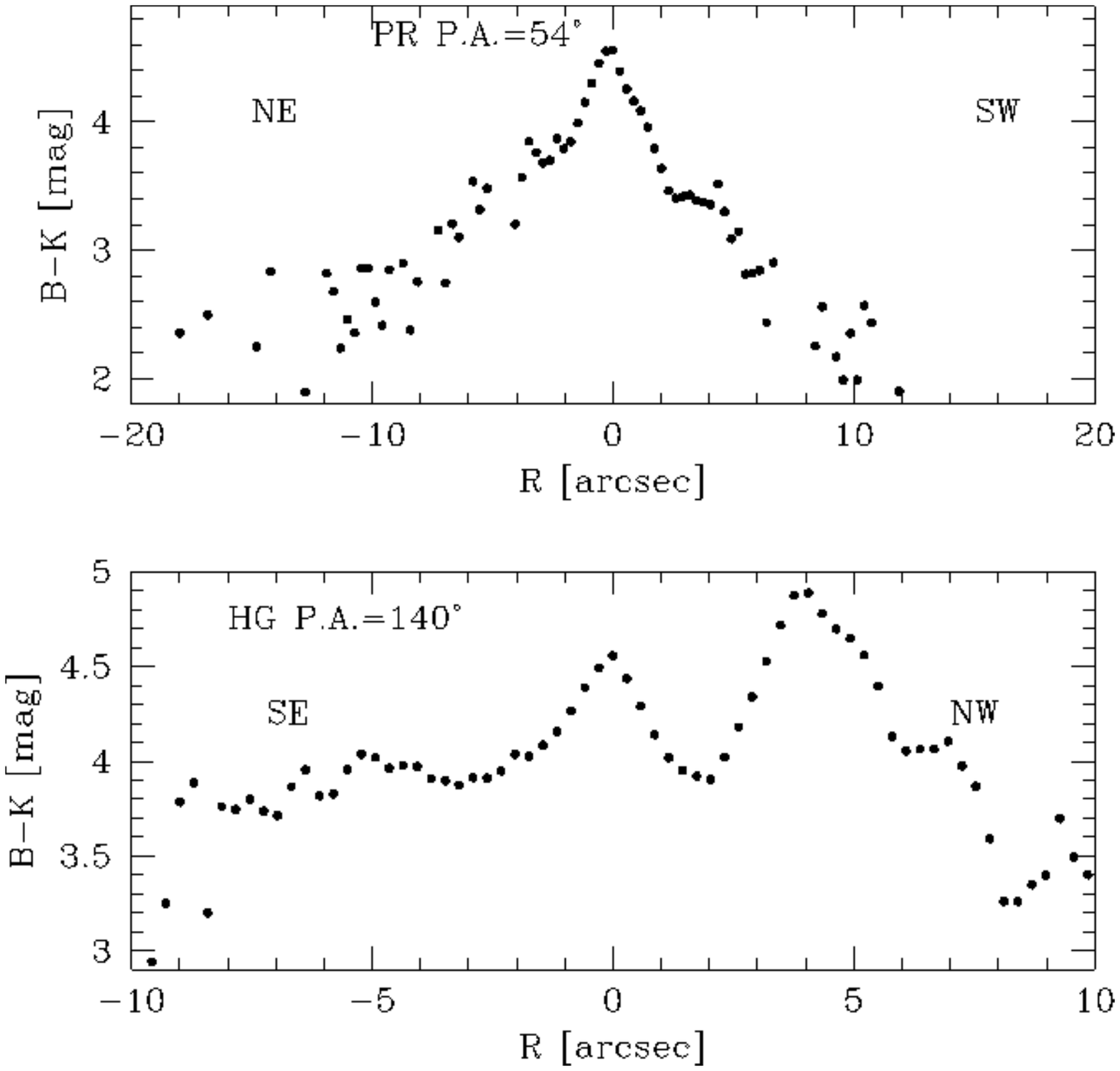}
\includegraphics[width=8cm]{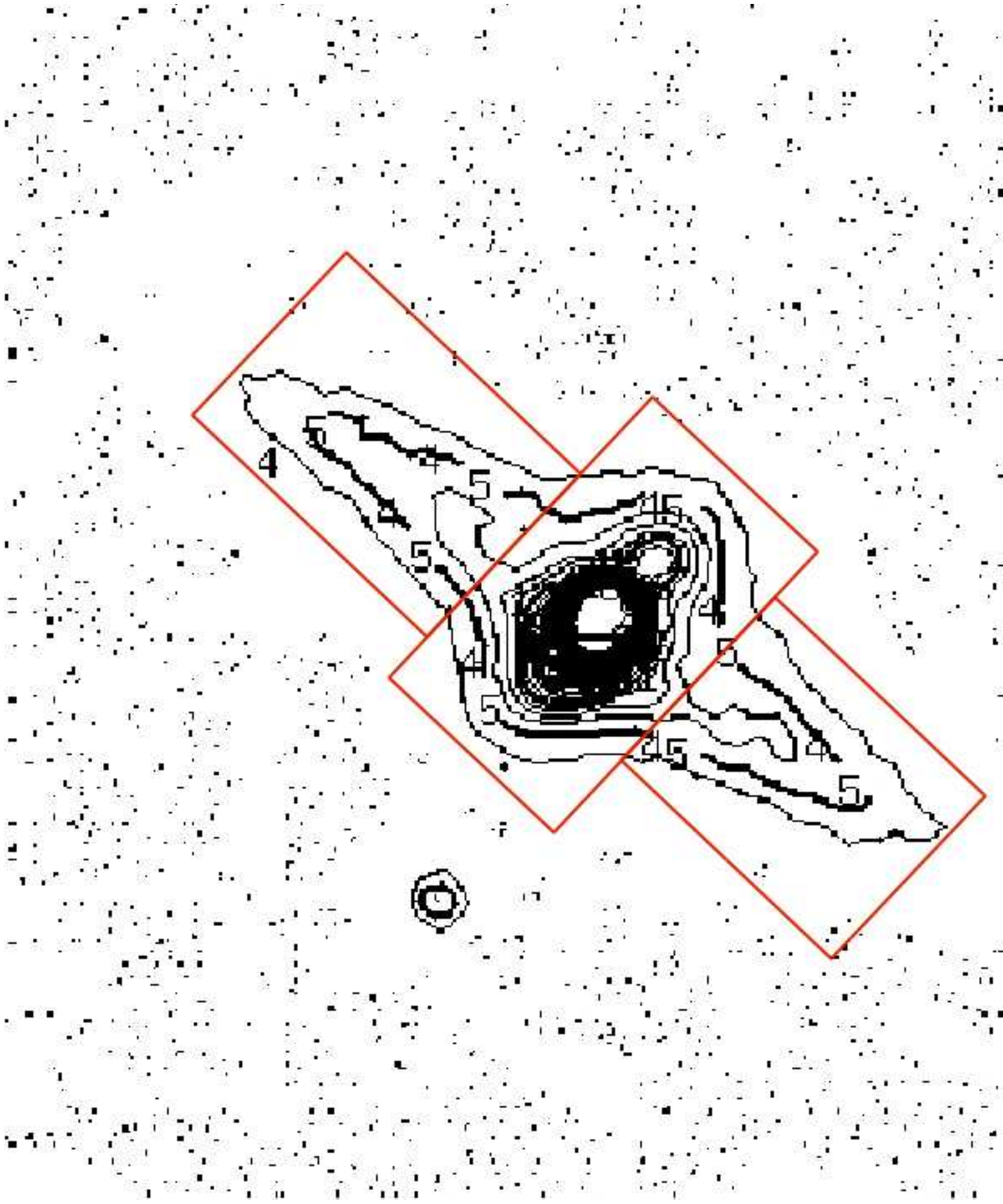}
\caption{Left panel - B-K color profiles along the HG major (bottom
  panel) and along the polar ring (top panel). The error bar ($\pm$
  0.03) is within the dimensions of data points. Right panel -
  Isophote contours of the B-band image of A0136-0801 with
  superimposed the three regions limiting the different areas where
  the integrated magnitudes have been computed. The North is up, while
  the east is on the left of the image.
} \label{B-K}
\end{figure*}

We have derived the B-K color profiles along the HG major axis, at
$P.A.=140^{\circ}$, and along the polar structure, at
$P.A.=54^{\circ}$, (see Fig.~\ref{B-K} left panel). The B-K color profiles show that nuclear
regions of the galaxy, for $R \le 3$~arcsec, have the reddest colors,
being $B-K=4.5\pm 0.03$~mag in the center. At larger radii, B-K
decreases. Along the HG major axis (see Fig.~\ref{B-K} bottom left
panel), on the SE side, B-K color profile is almost constant in the
range $3.7-4$~mag. On the NW side, for $3 \le R \le 6$~arcsec, the red
bump corresponds to the strong absorption by the dust in the polar
structure that passes in front of the HG (see also
Fig.~\ref{A0136JR}). The polar structure (see Fig.~\ref{B-K} top left
panel), for $R \ge 5$~arcsec is bluer than the HG, having
1.9$\le$B-K$\le$2.8~mag.

We derived the integrated magnitudes and J-K and B-K colors in three
areas\footnote{To do this, was used the IRAF task {\small POLYMARK} to
  define each area and the task {\small POLYPHOT} to derive the
  integrated magnitude.}, as shown in Fig.~\ref{B-K} (right panel): one
  including the central HG of A0136-0801 and two including the NE and
  the SW sides of the polar structure.  The three regions are traced
  on the B-band image, where the polar ring is much more luminous and
  extended with respect to the NIR images. Same polygons are adopted
  for all the other images (J and K bands), after they were registered
  and scaled to the B-band image. 
  Total magnitudes were corrected for the extinction
  within the Milky Way, by using the absorption coefficient in the B
  band $(A_{B})$ and the color excess E(B-V) derived from
  \citet{Schlegel98}. The absorption coefficients for the J and K
  bands are derived by adopting $R_{V}=A_{V}/E(B-V)=3.1$, and using
  the $A_{\lambda}$ curve from \citet{Car89}. The values of the
  absorption coefficients $A_{\lambda}$ adopted for A0136-0801 are
  $A_{B} = 0.115\ mag$, $A_{J} = 0.024\ mag$ and $A_{K} = 0.01\
  mag$. Moreover, in order to estimate the extinction law in the polar
  structure, we adopted the procedure described by \citet{Iod04}. In
  each band, we subtracted the unobscured part of the surface
  brightness profile (i.e. the northern side) from its obscured
  counterpart (the southern side), in order to obtain the ``absorption
  profile'' defined by
  
\begin{equation}
A_{\lambda} = -2.5 log \frac{I_{obs}(\lambda)}{I_{true}(\lambda)}
\end{equation} where $I_{obs}$ is the observed intensity in the polar structure and $I_{true}$ is the intensity relative to the starlight with no dust obscuration.
From this analysis we estimated the extinction coefficients $A_{B} = 0.24$~mag, $A_{J} = 0.2$~mag and $A_{K} = 0.17$~mag, used to correct the measured magnitudes.

The difference between the intrinsic absorption $A_{V}$ derived
  from the Balmer line ratios and the $A_{B}$, $A_{J}$ values derived
  from the photometry may be related with the different regions of the
  polar structure. While the $A_{V}$ is derived as an average quantity
  on both side of the major axis of the polar structure, $A_{B}$ and
  $A_{J}$ are computed locally in the obscured part of the HG where
  the polar ring passes in front of it.

Total magnitudes and colors and are listed in Table \ref{magnitudes}. The extinction corrected B-K integrated color
  for the HG derived in the area given above turns to be consistent with the same quantity estimated by
  the best fit 2D model of the light distribution for this component (see Sec. \ref{2D}), which is $B-K = 3.75 \pm 0.04$ mag.

\begin{table*}
\begin{minipage}{100mm}
\caption{\label{magnitudes}Integrated and absolute magnitudes and colors of different regions of A0136-0801 corrected for the MW and internal extinction.}
\begin{tabular}{lcccccccccc}
\hline\hline
Component& Region & $m_{B}$ & $m_{J} $ & $m_{K} $ & $M_{B}$ & $M_{J}$ & $M_{K}$ & B-K & J-K\\
         &        & $\pm 0.01$   & $\pm 0.02$   & $\pm 0.02$    &        & & & $\pm 0.03$& $\pm 0.04$\\
\hline
HG & center & 15.06 & 12.66 & 11.77 &-19.26 & -21.66 & -22.55 & 3.29 & 0.89\\
PR & NE & 16.39 &15.03 & 14.48 &-17.93&-19.29&-19.84& 1.91 & 0.55\\
PR & SW & 16.81 &15.69 & 15.15 &-17.51&-18.63&-19.17& 1.66 & 0.54\\
\hline
\end{tabular}
\end{minipage}
\end{table*}

\subsection{Stellar population analysis}

\begin{figure*}
\centering
\includegraphics[width=8.5cm]{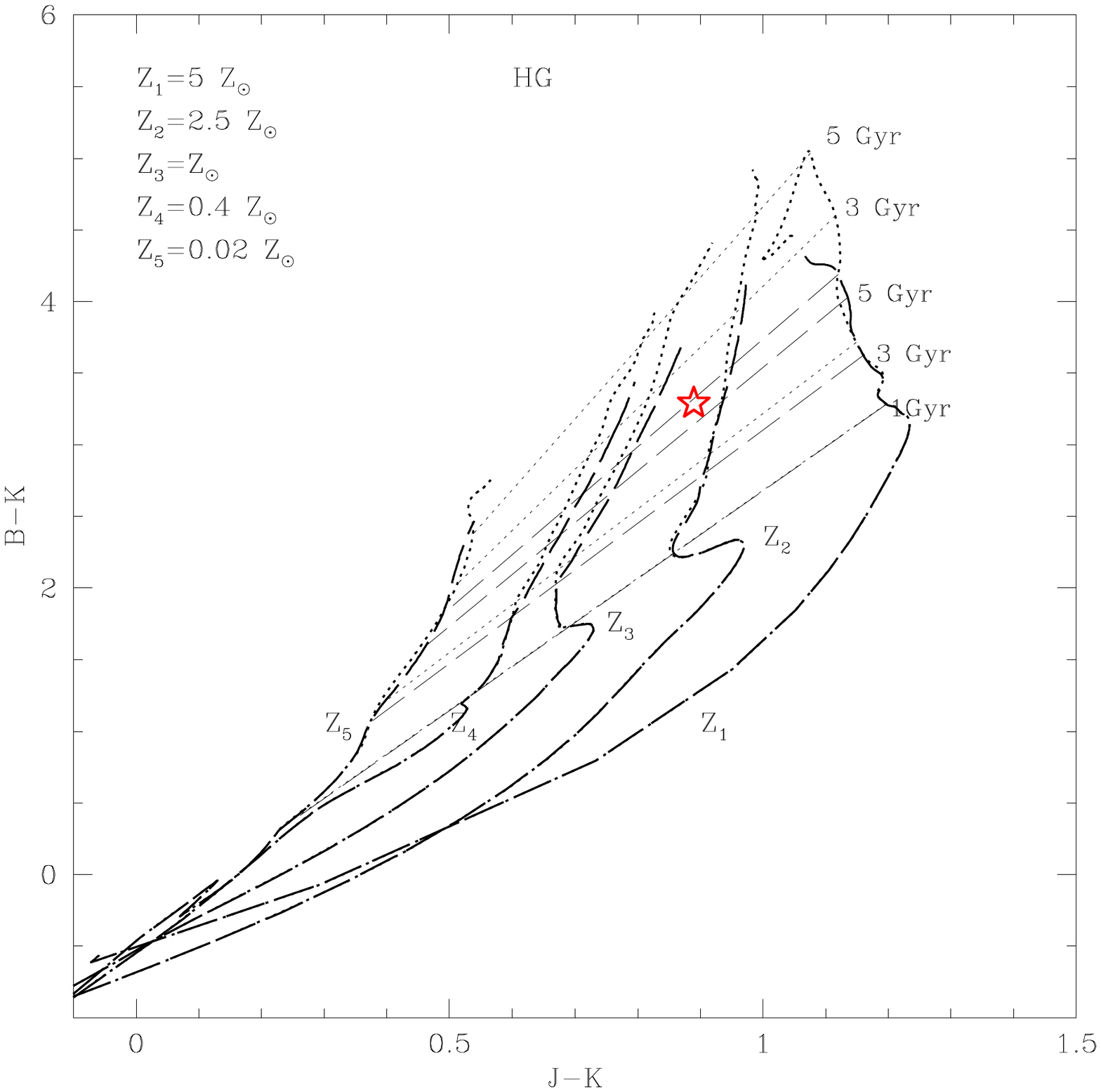}
\includegraphics[width=8.5cm]{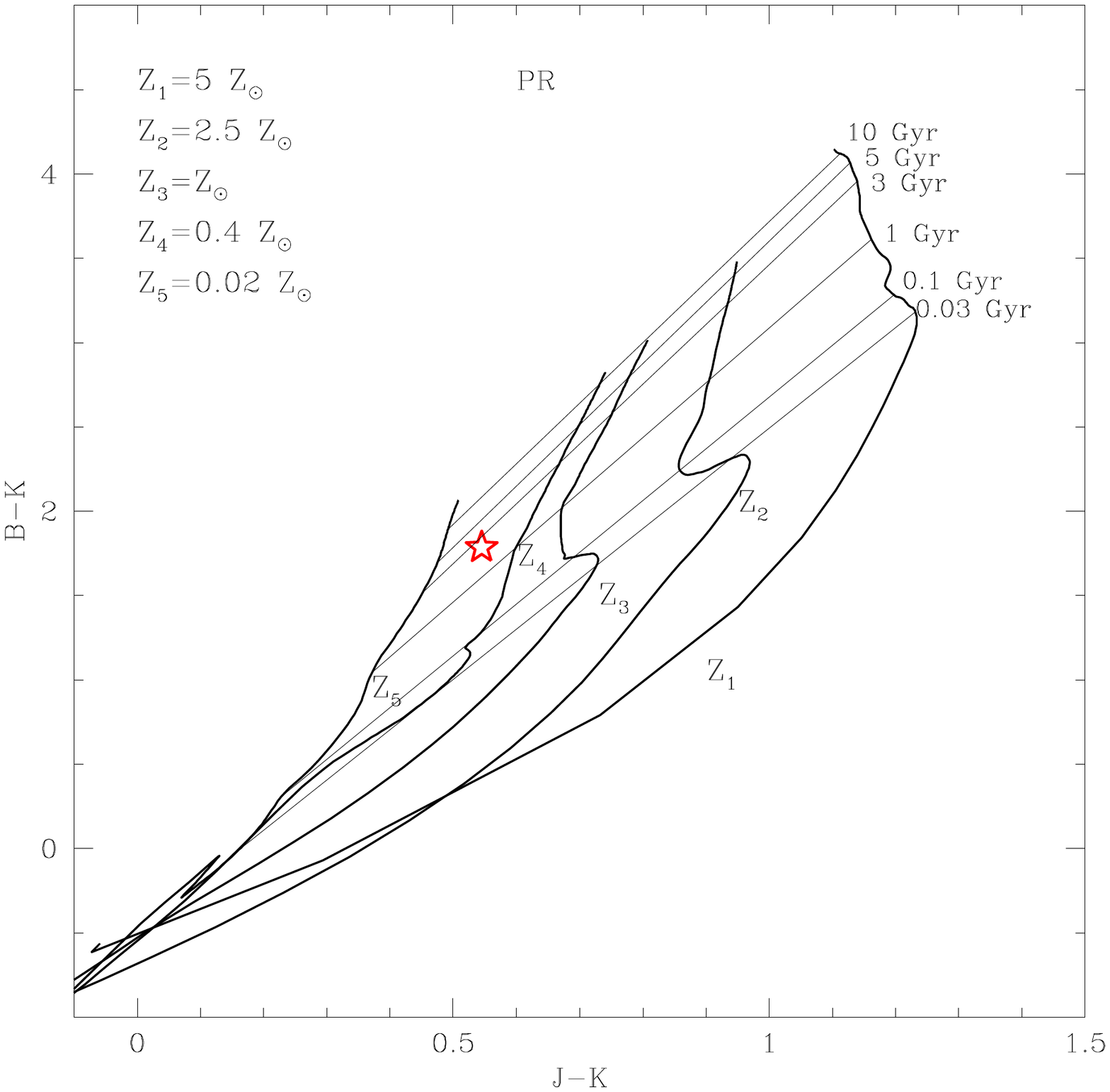}
\caption{B-K vs J-K color diagram for the HG (left panel) and polar
  ring (right panel) in A0136-0801. The red star corresponds to the
  average value, between NE and SW regions, for the polar structure in
  A0136-0801.  In both panels are drawn the evolutionary tracks
  derived by the stellar synthesis models optimized for the HG and PR
  in PRGs (see text for details). {\it Left panel} - For the HG, the heavy
  dotted lines corresponds to the SFR with a characteristic time scale
  $\tau$ = 1~Gyr, and the heavy dashed line to the SFR with $\tau$ = 7
  Gyrs. They are derived for different metallicities reported on the
  figure. The loci of constant age for the different tracks are
  indicated by the light dotted and light dashed lines. {\it Right panel} -
  For the polar structure, the heavy lines correspond to the models
  with a constant SFR computed for different metallicities. The loci
  of constant age for the different tracks are indicated by the light
  lines.}\label{age}
\end{figure*}

We aim to estimate the average (i.e. old plus the new bursts) stellar
population ages of the HG and polar structure in A0136-0801, by using
the integrated colors (optical vs NIR) measured for both components.
Following the same approach of \citet{Iod02a} and \citet{Iod02b}, we
derived the B-K and J-K integrated colors (see
Table~\ref{magnitudes}).  In the B-K versus J-K color diagram (shown
in Fig.~\ref{age}) are added the evolutionary tracks to reproduce the
integrated colors in the central component and in the ring-like
structures for a sample of PRGs, by using the stellar population
synthesis model GISSEL\footnote{\it Galaxies Isochrone Synthesis
  Spectral Evolution Library} \citep{Bru03}. The key input parameters
adopted for GISSEL are the Initial Mass Function (IMF), the Star
Formation Rate (SFR) and the metallicity. A detailed description of
them is given by \citet{Iod02b}.

For the central HG was adopted a star formation history with an
exponentially decreasing rate\footnote{It has the following analytical
  expression: $SFR(t) =1/\tau\ exp (-t/\tau)$, where the $\tau$
  parameter quantifies the ``time scale'' when the star formation was
  most efficient.}, that produces a reasonable fit of the photometric
properties of early-type galaxies in the local Universe.  For the
polar structure, which has usually bluer colors than the host galaxy
and presents HII regions, a constant SFR, typically used for local spiral galaxies, is used. The evolutionary tracks for each model were derived
for different metallicities ($Z_1= 5 Z_{\odot}$, $Z_2=2.5Z_{\odot}$, $Z_3=Z_{\odot}$, $Z_4=0.4Z_{\odot}$ and $Z_5=0.02Z_{\odot}$), which were assumed constant with age.

The analysis of the B-K versus J-K color diagram (see Fig.~\ref{age},
left panel) suggests that the stellar population in the HG in
A0136-0801 could be dated from 3 to 5~Gyrs, and has a metallicity in the range 
$Z_{\odot} \le Z \le 2.5 Z_{\odot}$. The polar ring  has a younger age (from 1 to 3~Gyr) and a sub-solar metallicity, in the range $0.02Z_{\odot} \le Z \le 0.4Z_{\odot}$, with respect to the HG. These values are comparable with those observed for the HGs and the PRs in the sample of PRGs in
\citet{Iod02a} and \citet{Iod02b}.

From the above models, we can derive the stellar mass-to-light ratio
(M/L) in the B band, for both HG and polar ring in A0136-0801, in
order to estimate the total baryonic mass (stars plus gas) for each
component.  For the central HG, the models predict an $M/L \simeq
2.98$~$M_{\odot}/L_{\odot}$. From the total magnitude in the B band for
the HG $M_B = -19.26$~mag (see Table~\ref{magnitudes}), the total
stellar mass is $M_b^{HG} \sim 2.3 \times\ 10^{10}$~M$_{\odot}$. Given
the absence of gas in this component, the baryonic mass coincides with
the stellar mass.
For the polar ring, the models predict an $M/L \simeq 2.18
$~M$_{\odot}/L_{\odot}$. This component contains a large amount of gas
($M_{HI} = 1.6 \times 10^9$~M$_{\odot}$ and $M_{H2} = 1.8 \times 10^9$~M$_{\odot}$, see Table~\ref{global}), thus, taking
into account the total magnitude in the B band, $M_B = -18.49$~mag
(see Table~\ref{magnitudes}),  the stellar mass is $M_{\ast}^{PR}
  \sim 8.4 \times\ 10^{9}$~M$_{\odot}$ and the total baryonic mass, stars plus gas,
is  $M_b^{PR} \sim 1.18 \times\ 10^{10}$~M$_{\odot}$. The importance of
these estimates on the formation history for A0136-0801 is discussed
in Sec.~\ref{conc}.

\section{Oxygen abundances and SFR in the polar structure of A0136-0801}\label{oxy}

We have derived the \emph{Oxygen abundance parameter} $R_{23} =
([OII]\lambda 3727 + [OIII]\lambda \lambda 4959 + 5007)/H \beta$
\citep{Pag79}, by following the procedure outlined by \citet{Spav10,Spav11} and \citet{Spav13}. 
According to \citet[][and references therein]{Spav13}, we used the \emph{Empirical method}
introduced by \cite{Pil01}. This allow us to estimate the oxygen abundance $12+log(O/H)$ and the
metallicity of the HII regions of the polar structure in A0136-0801. We found for this galaxy  $12+log(O/H) = 8.33 \pm 0.43$.

Assuming the oxygen abundance and metallicity of the Sun, $12 +
log(O/H)_{\odot} = 8.83 \pm 0.20= A_{\odot}$ and $Z_{\odot} = 0.02$
\citep{Gre98}, given that $Z \approx K Z_{\odot}$, where $K_{A0136} =
10^{[A_{A0136} - A_{\odot}]}$, we obtain a metallicity for the HII
regions in the polar structure of $Z \simeq\ 0.32 Z_{\odot}$.  This value turns to be consistent with the range of
metallicities derived for the stellar population ($0.02Z_{\odot} \le Z
\le 0.4Z_{\odot}$), given in Sec.~\ref{colors} (see also
Fig.~\ref{age}).

\begin{figure*}
\centering
\includegraphics[width=12cm]{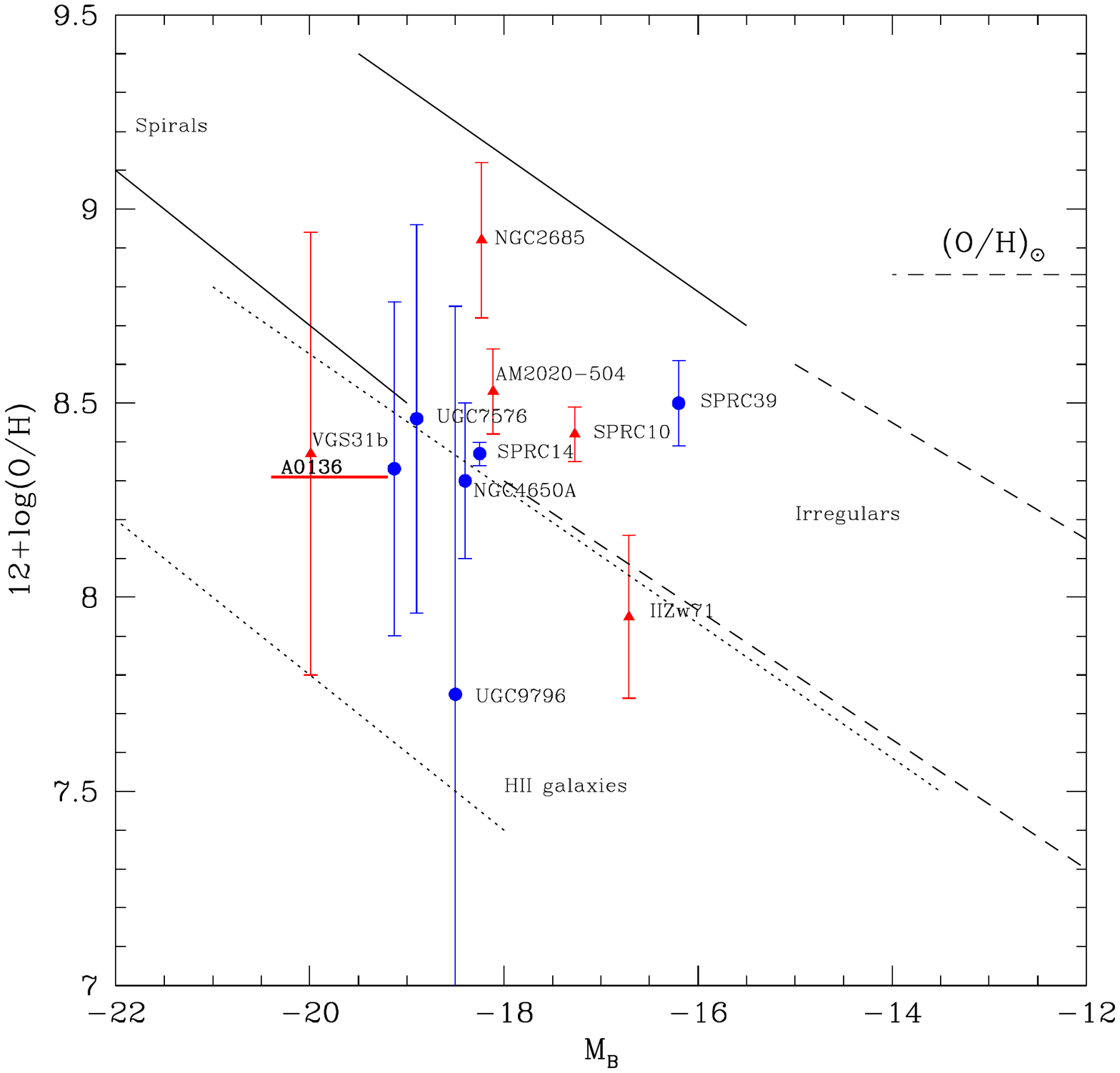}
\caption{Oxygen abundance vs absolute blue magnitude for A0136-0801
  (from this work) and for other PRGs: VGS31b \citep{Spav13}, NGC4650A
  \citep{Spav10}, IIZw71 \citep{Per09}, NGC2685 \citep{Esk97},
  AM2020-504 \citep{Frei12}, UGC7576 and UGC9796 \citep{Spav11},
  SPRC10-14-39 \citep{Moi13}. Blue circles represent wide PRGs, while
  red triangles are for narrow PRGs. The sample of late-type disk
    galaxies are by \citet{Kob99}: Spirals are in the region marked
    with continuous lines, Irregulars are between dashed lines and HII
    galaxies are between dotted lines. The horizontal dashed line
    indicates the solar oxygen abundance.} \label{conf}
\end{figure*}

In Fig.~\ref{conf} we compare the mean value of the oxygen abundance
along the polar structure of A0136-0801, with those measured for
different late-type galaxies by \cite{Kob99}\footnote{The absolute
  blue magnitude for the objects in the sample of Figure 4 in
  \citet{Kob99} are converted by using $H_{0} =75$ km/s/Mpc.}
(spirals, Irregulars and HII galaxies) and several PRGs, as a function
of the total luminosity.  The metallicity of the polar structure in
A0136-0801 is comparable with that observed for both wide and narrow
PRGs, which show a sub-solar values with respect to the mean
metallicity of spiral galaxies with similar total luminosity. On
average, taking into account the mean error on the oxygen abundances,
PRGs are found in the region of low luminosity spirals or/and bright
irregulars.

From the $H \alpha$ luminosity derived from emission lines with $S/N >
10$, using the expression given by \cite{Ken98}, we estimate the Star
Formation Rate (SFR) for the polar structure of A0136-0801, this is
$SFR = 7.9 \times 10^{-42} \times L(H \alpha)$. From the average value
of $L(H \alpha) \simeq 1.11 \times 10^{38}$ erg/s we have obtained an
average $SFR \sim 9 \times\ 10^{-4} M_{\odot}/yr$. This value of
  SFR has to be considered as a lower limit, since the long slit data
  give only lower limits to the $H \alpha$ luminosity, given that the
  slit usually covers only small area of the polar structure. A global
  estimate of the SFR in this component can be inferred from the Kron
  UV flux for A0136-801 from NED. We derive $L(UV)=3.8
  \times\ 10^{26}$ erg/s, with E(B-V) = 0.026 \citep{Seibert12}, and from the Kennicutt law \citep{Ken98}, we
  obtain $SFR = 1.4 \times\ 10^{-28} \times\ L (UV) \sim\ 0.05
  M_{\odot}/yr$, which is a more reliable estimate for the SFR in the
  polar structure of A0136-0801.

As done in similar studies for other PRGs \citep{Spav10,Spav11,Spav13}, we checked whether the present SFR, 
and even 2 and 3 times higher can give the inferred metallicity of $Z=0.32 Z_{\odot}$. 
To this aim, we adopted the linearly declining SFR, 
$SFR(t) =2 M_{\star}\tau^{-1}exp[1-(t/\tau)]$, where $t$ is the lookback time \citep{Bru03}.
By using the mass-metallicity relation derived by \cite{Tre04}, we found that for A0136-0801
the expected metallicity is  in the range $0.1 Z_{\odot} \le Z \le
0.66 Z_{\odot}$. 
The implications of this result will be discussed in detail in Sec.~\ref{conc}.

\section{Discussion and conclusions}\label{conc}

We have carried out a detailed photometric study of the polar ring
galaxy A0136-0801, based on new NIR, in the J and K bands, and
optical, B and R bands, observations. Moreover, by using new long slit
spectra, we have studied the chemical abundances in the polar
structure of this galaxy. In this section we summarise the main
results and discuss their implications on the formation history of
A0136-0801.

\subsection{Summary of the main results}

The morphological analysis performed in this work shows that
A0136-0801 has two main components, the central spheroid (the HG),
with a bright nucleus, and a wide polar structure (see Fig.~\ref{A0136JR}).

\begin{enumerate}

\item The HG dominates the light in the NIR images and its structure
  can be well traced at all radii since the dust absorption due to the
  polar ring is negligible (see Fig.~\ref{A0136JR}, right panel). In
  the optical images, the polar ring is much more luminous and it is
  at least two times more extended than the HG (see
  Fig.~\ref{A0136JR}, left panel).

\item The 2D models of the light distribution performed in both B and
  K bands have shown that the HG is characterised by a very
  concentrated bulge-like structure, with an effective radius of
  1.12~arcsec ($\sim0.34$~kpc) and and exponential disk, with a scale
  length of about 3~arcscec ($\sim0.9$~kpc), see
  Table~\ref{galfit}. The best fit to the light distribution of the
  polar structure has been obtained in the B band, with an exponential
  law, having a scale length of 13.3~arcsec ($\sim4$~kpc), which is
  four times larger than the scale length of the disk in the HG.  The
  residual map, both in the dust-free K-band and in the B-band images,
  have revealed the existence of a nuclear sub-structure having an
  ``s-shape'' (see right panels of Fig.~\ref{modK}),
  elongated towards the polar direction.

\item From the B-K color distribution, we found that the HG
  has redder colors with respect to the polar structure (see
  Fig.~\ref{B-K}). In particular, the B-K color profiles show that
  nuclear regions of the galaxy, for $R \le 3$~arcsec, have the
  reddest colors, being $B-K=4.5\pm 0.03$~mag in the center. On
  average, the HG has a $B-K \sim 3.7-4$~mag. The polar structure (see
  Fig.~\ref{B-K} top left panel), for $R \ge 5$~arcsec is bluer than
  the HG, having B-K$\le 3$~mag.

\item By using optical versus NIR colors (B-K vs J-K) and the stellar
  population synthesis models we estimated the age and metallicity of
  the stellar population in both components of A0136-0801. The
  constraints on the age and metallicity of the stellar population in
  the HG is within the range $3-5$~Gyrs, and $Z_{\odot} \le Z \le 2.5
  Z_{\odot}$. The polar ring has a younger age (from 1 to 3~Gyr) and a
  sub-solar metallicity in the range $0.02Z_{\odot} \le Z \le
  0.4Z_{\odot}$ (see Fig.~\ref{age}). From the above models we
  estimate the total baryonic mass, for both HG and polar ring in
  A0136-0801 (see Sec.~\ref{colors}), values are listed in
  Table~\ref{scenarios}.

\item We derived the oxygen abundance and the metallicity of the HII
  regions of the polar structure in A0136-0801.  We found $12+log(O/H)
  = 8.33 \pm 0.43$ and $Z \simeq\ 0.32 Z_{\odot}$ (see
  Sec.~\ref{oxy}). This value turns to be consistent with the range of
  metallicities derived for the stellar population ($0.02Z_{\odot} \le
  Z \le 0.4Z_{\odot}$).

\end{enumerate}

\subsection{Comparison of A0136-0801 with other PRGs}

We discuss the observed properties of A0136-0801 with known PRG
systems.  The morphology of A0136-0801 is very similar to those of
other PRGs with a wide polar structure, like NGC~4650A, UGC~7576 and
UGC~9796. A detailed photometric analysis have shown that in UGC~7576
and UGC~9796 \citep{Res97,Godinez07} this component is a ring, while
in NGC~4650A it is a disk \citep{Iod02}. As discussed above, the
residual map obtained for A0136-0801 by the 2D model of the light
distribution, shows a nuclear ``s-shaped'' structure, elongated
towards the polar direction, which could be a sign of the polar
structure reaching the HG center, as that observed in NGC~4650A
\citep{Iod02, Gal02}. Contrary to what the photometric analysis
suggests, the integrated colors and metallicity of the stellar
populations in the polar structure of A0136-0801 are quite different
from those observed for the same component in NGC~4650A \citep{Iod02,
  Spav10}, being older and with higher metallicity.  In
Table~\ref{scenarios} are also included the key physical parameters
available for the wide PRGs mentioned above and for the narrow PRG
AM2020-504.

All PRGs are characterised by a large amount of HI gas. The HI mass
measured in A0136-0801 is of the same order of magnitude of those
found for other PRGs ($\sim 10^{9}$~M$_{\odot}$), except NGC~4650A
that is the PRGs with the highest HI mass ($\sim
10^{10}$~M$_{\odot}$).  As found in A0136-0801 (see
Sec.~\ref{colors}), the large baryonic mass in the HG with respect to
that in the PR is an observed feature common to all PRGs, except for
NGC~4650A, where the polar disk is twice as massive as the central
spheroid. The HG in all wide PRGs, including A0136-0801, is supported
by rotation, with $v/ \sigma > 1$, while the HG in the narrow PRG
AM2020-504 is dominated by random motions.  Overall, the metallicity
of the polar structure in A0136-0801 is comparable with that observed
for most of the wide and narrow PRGs, which show lower,
i.e. sub-solar, values with respect to the metallicity in spiral
galaxies of similar total luminosity (see Fig.~\ref{conf}). In
particular, the metallicity in the polar structure of A0136-0801 is
comparable with that measured for the wide PRG UGC~7576, while it is
two times larger with respect to those estimated for UGC~9796. The
narrow PRG AM2020-504 is among the PRG with the highest value of $Z$,
which is close to the solar metallicity.  As already pointed out by
\citet{Iod2014}, the differences in these key parameters are probably
related to different formation processes.

\subsection{Implications on the formation history of A0136-0801}

In order to address the most reliable formation scenario for
A0136-0801, below we compare and discuss how the observed properties
for this galaxy, outlined in the present work, could compare with the
predictions of different formation mechanisms for PRGs.  In
particular, we focus on the structure, i.e. the presence of a wide
ring around a spheroidal object, on colors and ages for both HG and
PR, on the observed kinematics and the gas content and metallicity in
the polar structure.

As reviewed into Sec.~\ref{intro}, a PRG could form by the i) the
disruption of a dwarf companion galaxy or of a gas-rich satellite, in
the potential of an early-type system, or by the tidal accretion of
gas stripped from the outskirts of a disk galaxy, ii) a dissipative
merging of two disk galaxies, or iii) the accretion of cold gas from
cosmic web filaments.

The key physical parameters that can discriminate between the
different formation scenarios are {\it i)} the total baryonic mass
(stars plus gas) observed in the polar structure with respect to the
central spheroid, {\it ii)} the kinematics along both the equatorial
and meridian planes, {\it iii)} the metallicity and SFR in the polar
structure. For A0136-0801, all the above parameters are now available,
obtained by the analysis performed in this work (i.e. the total
baryonic mass and metallicity) and from literature (gas content and
kinematics), they are listed in Table~\ref{scenarios}.  The baryonic
mass in the polar structure of A0136-0801 is less than that in the HG,
being the ratio $M_{b}^{HG}/M_{b}^{PR} \sim 1.95$. The central HG is
supported by rotation, with $v/ \sigma \sim 2.2$. The polar structure
has a sub-solar metallicity ($Z \simeq\ 0.32 Z_{\odot}$).

The tidal accretion scenario, in which the polar structure forms
through the gas stripped from a gas-rich donor galaxy, in a
high-inclined orbital configuration \citep{Bou03}, is able to produce
a wide polar ring as observed in A0136-0801. During this kind of
interaction, the central HG does not change its original morphology
and kinematics.  So, as observed in A0136-0801, the remnant is an
S0-like system, supported by rotation, with a polar structure.  In
this framework, in the field around the newly formed PRG the gas-rich
donor galaxy should be still present.
In the case of A0136-0801, inside a radius of about five times its
diameter, as suggested by \citet{Broc97}, there is a late-type galaxy
PGC~6186 at a comparable redshift (see Fig.~\ref{field}). This object
is very poorly studied and only the apparent magnitude in the B band
is given, which is 15.20 mag (see NED). There is no information on the
gas content. Anyway, since this object has a comparable luminosity to
that observed for A0136-0801, and it is classified as spiral, we expect
that it has at the least the same, or even larger, baryonic mass of
A0136-0801 (see Table~\ref{scenarios}). Thus, taking also into account that the
value of $Z\simeq\ 0.32 Z_{\odot}$ derived for A0136-0801,
is consistent with the metallicity of the very outer regions of bright
spiral galaxies, which is in the range $0.2 Z_{\odot} \leq Z \leq 1.1
Z_{\odot}$ \citep{Bre09}, PGC~6186 could be considered a possibile
donor galaxy.

The gradual disruption of a dwarf satellite galaxy can be excluded as
possible formation process for A0136-0801, because the baryonic mass
estimated in the polar structure ($1.18\times\ 10^{10} M_{\odot}$, see
Table~\ref{scenarios}) is at least 3 order of magnitude larger than
the typical mass observed in dwarf galaxies \citep[$10^{3} M_{\odot}
  \leq M \leq 10^{7} M_{\odot}$][]{Saw11}. On the hand, if in the
past A0136-0801 had a gas-rich satellite, with a mass
of $\sim 10^9 M_{\odot}$ \citep{Kunkel96}, like the Large Magellanic Cloud, it could be
disrupted by the potential of the massive HG to form the polar
structure.

In the merging scenario, the morphology and kinematics of the merger
remnants depend on the initial orbital parameters and the
initial mass ratio of the two merging galaxies \citep{Bou05}. For A0136-0801,
this scenario is ruled out because, according to simulations
(e.g. \citealt{Bou05}), a high mass ratio of the two merging galaxies
is required to form a massive and extended polar ring as observed in
A0136-0801. This would convert the intruder into an elliptical-like,
not rotationally supported, stellar system. This prediction is in
contrasts with the observed kinematics for the HG in A0136-0801.

Finally, we discuss whether the cold accretion scenario can reliable
account for the observed properties of A0136-0801. This scenario
predicts the formation of wide disk-like structures, characterised by
a low, sub-solar metallicity. 
The open issue of this scenario concerns the nature of the polar structure: the simulations
predict the formation of a polar disk, rather than annulus, around a
puffed up disk \citep{Mac06, Bro08}. The residual map, obtained for A0136-0801
by the the 2D model of the light distribution, have revealed the
existence of a nuclear sub-structure having an ``s-shape'' (see right
panels of Fig.~\ref{modK}), elongated towards the
polar direction. If this feature is linked to the polar structure that
reaches the galaxy center, this could be an hint for a polar disk in
A0136-0801, but a definitive conclusion on the nature of the polar component in A0136-0801 cannot
be reached on the basis of the photometry done. The kinematics mapping the
inner regions\footnote{The kinematics published by \citet{Sch83} is
  relative only to the outer arms of the polar structure and along the
  major axis of the HG. } is also necessary to unveil the structure of this galaxy.
Thus, the available data for A0136-0801 cannot allow us to discriminate between a ring or a
disk. 
Anyway, even if the polar structure in this galaxy was a disk,
it is characterised by an higher metallicity ($Z \simeq\ 0.32 Z_{\odot}$) than that predicted by the cold accretion scenario,
which is $Z \leq 0.2 Z_{\odot}$ \citep{Snaith12}. Moreover, the
metallicity estimated by using element abundances falls in the range
of those expected by the SFR, which is $0.1 Z_{\odot} \le Z \le
0.66 Z_{\odot}$ (Sec.~\ref{oxy}). Given that these values could further
increase the metallicity of $\sim 0.01$ after 1 Gyr, we can rule out
the cold accretion scenario for A0136-0801.

To conclude, even if the whole morphology of the polar ring galaxy
A0136-0801 is very similar to that observed for NGC~4650A, both are
classified as ``wide PRGs'', the two system have a different formation
history: the polar disk in NGC~4650A is reasonably formed by the cold
accretion of gas along cosmic web filaments \citep{Spav10}, while for
the PRG A0136-0801 the tidal accretion of material (gas and stars)
from the outskirts of a donor galaxy, as well as the tidal
  disruption of a gas-rich satellite, are the most viable formation
processes.


\begin{table*}
\begin{minipage}{140mm}
\caption{Discriminating parameters between different formation scenarios}   
\label{scenarios}      
\begin{tabular}{l c c c c c c c c  c}        
\hline\hline                 
PRG & $M_{b}^{HG}$ & $M_{b}^{PR}$ & $V_{eq}$ & $V_{eq}/V_{p}$ & $\sigma_{0}$ & $M_{b}^{HG}/M_{b}^{PR}$ & $M_{HI}$& Z&  Ref.\\    
        & $10^{9} M_{\odot}$ & $10^{9} M_{\odot}$ & km/s & & km/s & & $10^{9} M_{\odot}$ & $(Z_{\odot})$ & \\
(1)   & (2) & (3) & (4) & (5) & (6) & (7) & (8) & (9) & (10) \\
\hline                        
   {\it A0136-0801} & 23 & 11.8   &  145 &  0.9   &  67 &  1.95& 1.6 & 0.32 & {\em a, b}\\
   \hline
   UGC7576 & 7.86  & 2.88 &  212 & 0.96  & 116 & 2.73 & 2.7 & 0.4 & {\em c}\\      
   UGC9796  & 10.0 & 3.05 &  157 & 1.08  &  73 &  3.28 & 2.6 & 0.1 & {\em c}\\
   NGC4650A  & 5    & 12    &  90  &  0.75  &  70 &  0.42 & 8.0 & 0.2 & {\em d, e}\\
   AM2020-504 & 1.16   & 1.04     &  120 &  0.48  & 260 &  1.1 & 2.7 & 0.5-1 & {\em f, g}\\
\hline                                   
\end{tabular}
{\em Col.1}: PRG {\em Col.2}: Total baryonic mass in the central
galaxy of the PRG. {\em Col.3}: Total baryonic mass in the polar
structure.  {\em Col.4}: Maximum rotation velocity along the major
axis (equatorial plane) of the HG. {\em Col.5}: Ratio between the
maximum rotation velocity along the equatorial and polar
directions. {\em Col.6}: Central velocity dispersion of stars in the HG. {\em
  Col.7}: Ratio between the baryonic mass of the HG and polar
ring. {\em Col.8}: Mass of the HI. {\em Col.9}: Metallicity of the HII
regions in the polar structure. {\em Col.10}: References containing
the values reported in the table.\\ {\em a, b}: \citet{Sch83},
\citet{vanDriel00}, this work\\ {\em c}: \citet{Spav11} and references
therein\\ {\em d, e}: \citet{Iod02}; \citet{Iod06} and references
therein\\ {\em f, g}: \citet{Arn1993}; \citet{Van02}; \citet{Frei12}\\
\end{minipage}
\end{table*}

\section*{Acknowledgements}
We are very grateful to the anonymous referee for his/her comments and
suggestions, which helped us to improve and clarify our work.
M.S. wish to thank T. Pursimo for the support given during the data
acquisition at NOT and D. Bettoni for many useful discussions and
suggestions.  E.I. wish to thank E. Pompei for the support given
during the data acquisition at NTT. This work is based on observations
made with ESO Telescopes at the Paranal Observatories under programme
ID $<$ 70.B - 0253(A) $>$ and $<$ 74.B - 0626(A) $>$. The authors wish
to thank A. Watson for providing optical images analyzed in this
work. The data presented here were obtained [in part] with ALFOSC,
which is provided by the Instituto de Astrofisica de Andalucia (IAA)
under a joint agreement with the University of Copenhagen and NOTSA.

\bibliography{bibliografia}

\bsp

\label{lastpage}

\end{document}